\documentclass{article}
\usepackage{graphicx} 
\usepackage[utf8]{inputenc}
\usepackage{algorithm}
\usepackage{algpseudocode}
\usepackage{natbib}
\usepackage{amsmath}
\usepackage{hyperref}
\usepackage{multirow}
\usepackage{multicol}
\usepackage{authblk}
\usepackage{xcolor}
\title{On a Semiparametric Stochastic Volatility Model}
\date{}
\author[*]{Yudong Feng}
\author[*]{Ashis Gangopadhyay}
\affil[*]{\textit{Department of Mathematics and Statistics, Boston University, 665 Commonwealth Ave, Boston, MA 02215, USA.}}
\affil[ ]{E-mail: ydfeng@bu.edu, ag@bu.edu}
\begin{document}

\maketitle

\begin{abstract}

This paper presents a novel approach to stochastic volatility (SV) modeling by utilizing nonparametric techniques that enhance our ability to capture the volatility of financial time series data, with a particular emphasis on the non-Gaussian behavior of asset return distributions. Although traditional parametric SV models can be useful, they often suffer from restrictive assumptions regarding errors, which may inadequately represent extreme values and tail behavior in financial returns. To address these limitations, we propose two semiparametric SV models that use data to better approximate error distributions. To facilitate the computation of model parameters, we developed a Markov Chain Monte Carlo (MCMC) method for estimating model parameters and volatility dynamics. Simulations and empirical tests on S$\&$P 500 data indicate that nonparametric models can minimize bias and variance in volatility estimation, providing a more accurate reflection of market expectations about volatility. This methodology serves as a promising alternative to conventional parametric models, improving precision in financial risk assessment and deepening our understanding of the volatility dynamics of financial returns.


\end{abstract}

\section{Introduction}

In the realm of financial time series analysis, the volatility of asset returns occupies a central position. The significance of volatility modeling extends across a broad spectrum of practical financial applications, encompassing asset pricing, portfolio optimization, and risk management, among others. Given its pivotal role over three decades, a significant effort has been made in financial literature to research and develop methods for estimating volatility. Prominent among these modeling techniques are the Auto-regressive Conditional Heteroskedasticity (ARCH) model, pioneered by \cite{ARCH}, the Generalized ARCH (GARCH) model by \cite{GARCH}, and the Stochastic Volatility (SV) Model, which was conceptualized by \cite{Taylor}. In particular, the Taylor style model of returns $y_1, y_2,..., y_N$ is
\begin{equation}
    \begin{aligned}
        y_t & =\sqrt{h_t}u_t,\  t = 1,...,N \\
        \ln h_t & =\alpha+\delta\ln h_{t-1}+\sigma_\nu\nu_t,\ t = 2,...,N
    \end{aligned}
    \label{eq: model}
\end{equation}
where $u_t,\nu_t\sim N(0,1)$ for $t = 1,...,N$ and $corr(u_t,\nu_{t'}) = 0$ for all $t$ and $t'$. 

Further studies on similar model frameworks have been conducted by various researchers, including 
\cite{Hull_White}, \cite{Chesney_Scott}, \cite{Taylor}, \cite{Jacquier}, \cite{Jacquier2002}, and \cite{Shephard}. Their works delve deeper into the methods for both parametric and semi-parametric estimation of stochastic volatility models. \cite{Harvey1994} introduced an asymmetric stochastic volatility model that incorporates the leverage effect, utilizing a quasi-maximum likelihood estimation approach within a state-space framework, which facilitates the analysis of financial time series displaying asymmetric volatility patterns. \cite{Carter_Kohn} developed a Gibbs sampling technique for examining state-space models, particularly effective in cases where errors are governed by mixture distributions and coefficients evolve over time. \cite{Kim1998} suggested a Bayesian approach using MCMC methods for stochastic volatility models, highlighting the advantages of SV models over ARCH models in representing the volatility dynamics of financial time series while introducing methods for filtering, diagnostics, and model selection. In addition, \cite{Omori2007} broadened the Bayesian analysis of SV models to account for leverage effects and presented a robust MCMC-based technique that employs a mixture approximation for the joint innovation distributions. 

In financial econometrics, Taylor's Stochastic Volatility Model (SVM) offers the notable advantage of accommodating time-varying volatility. However, as critiqued by \cite{Durham2006}, the model is hindered by its inability to encapsulate the commonly observed heavy-tailed behavior in the conditional distribution of returns. This shortcoming stems from its Gaussian assumption for the error terms. A central objective within financial econometrics is examining the returns distribution from financial markets, which yields insights into the mechanisms propelling financial dynamics. \cite{Mandelbrot1963} on cotton price fluctuations revealed pronounced heavy-tailedness compared to normal distributions, prompting the exploration of stable distributions as a superior alternative to conventional Gaussian distribution models. 

To tackle this issue, alternative parametric heavy-tailed distributions like the Student's t distribution with low degrees of freedom or the Generalized Error Distribution (GED) are options to explore. Nonetheless, as highlighted in the articles by 
\cite{efficiency_semi_para}, \cite{One-step}, when the actual error distribution is not known, these parametric assumptions frequently fall short. They often fail to capture the structural behavior of financial data accurately. 

One possible solution to this constraint is to utilize a non-parametric distribution for the return error term. This approach would offer increased flexibility in capturing the empirical characteristics of financial data, effectively allowing the data to ``speak for itself'' Similar concepts have been explored in prior research by \cite{One-step} and \cite{efficiency_semi_para} concerning GARCH models. The authors introduced a novel method for nonparametrically estimating the error distribution, which led to the development of a pseudo-likelihood function for the model. The resulting volatility derives from the GARCH parameters estimated via this likelihood.

In this paper, we propose a similar semiparametric approach to estimating volatility based on the SVM. However, the problem is significantly more challenging compared to the semiparametric formulation of GARCH. First, the SVM model introduced in Equation \ref{eq: model} requires the specification of the error distributions for both the observation equation $u_t$ and the volatility equation $\nu_t$.  Therefore, the semiparametric SVM introduced in this paper involves a novel methodology to utilize the latent features of the model to develop nonparametric estimates of the error terms. Second, the resulting pseudo-likelihood poses a significant optimization problem requiring a computational solution. Therefore, in the pursuit of constructing a semiparametric model, the deployment of numerical methods becomes vital. This paper discusses a computational solution based on the Markov Chain Monte Carlo (MCMC) technique for SV models rooted in the pioneering work by \cite{Jacquier2004}. This foundational work introduces Bayesian inference mechanisms for conventional parametric SVM discussed in Equation \ref{eq: model}.

The approach utilized in this paper involves a two-step process. We begin by fitting an initial parametric model; the derived volatility estimate is denoted by $\hat{h}$. Subsequently, the initial residual $\hat{u_t}$ can be defined as $y_t / \hat{h}$, which can be used to develop a kernel density estimate $f$, the pdf of the error term $u_t$. Based on the resulting estimate, $\hat{f}$, along with a parametric assumption on the volatility error term $\nu_t$, it becomes feasible to fit the semiparametric SVM model employing analogous sampling techniques. 

However, the ultimate objective of the current work is to develop a full semiparametric framework of the SVM. Towards this goal, we propose an approach to jointly estimate the density of the error terms $u_t$ and $\nu_t$. This is done by recursively approximating the residual of the observation equation as described earlier, along with the residual term from the volatility equation, denoted as $w_t$, i.e., $w_t=\frac{\ln h_t - u_t}{\sigma}$, which is utilized to estimate $g$, the pdf of the volatility noise $\nu_t$. 

This methodology parallels our treatment of the initial residual, wherein we employ kernel density estimation to approximate the error distribution for both residuals. Consequently, in addition to $f$, the function $g$  is utilized as the standardized version of the kernel density estimate of $\frac{w_t - mean(w_t)}{sd(w_t)}$. Embracing this semiparametric paradigm, the comprehensive model can be effectively operationalized within a Bayesian framework, employing Markov Chain Monte Carlo (MCMC) techniques to facilitate its implementation.

The advantage of the semiparametric model can be proved by comparing it to its parametric counterpart. This comparative analysis was facilitated using simulated data sets featuring both normally distributed innovations and those with heavy-tailed innovations, such as the Student's t distribution and the Generalized Error Distribution (GED). Additionally, empirical data from the S$\&$P 500 was also utilized in this endeavor. The resultant findings elucidate that the non-parametric model boasts a reduced variance coupled with an amplified precision in parameter estimation.

The rest of the paper is organized as follows: Section 2 introduces MCMC sampling in the context of the MCMC sampling scheme utilized in the paper; sections 3 and 4 contain the key contribution of the paper by proposing the pseudo-likelihood of the semiparametric SVM  and a Bayesian MCMC methodology to estimate the model parameters and resulting volatility. Sections 3 and 4 illustrate the proposed model's performance via extensive simulation and an analysis of the S\&P 500 dataset. The paper concludes with a discussion of the implications of the results.

\section{Semiparametric Estimation in Stochastic Volatility Model}
 
 This section will introduce the concept and related computational tools for semiparametric SVM.  However, in the next subsection, we will first review a Bayesian framework for SVM and the related computational tools.
 
\subsection{A Bayesian framework of SVM}

The Taylor-style stochastic volatility model is 
\begin{equation}
    y_t=\sqrt{h_t}u_t,\  t = 1,...,N 
\end{equation}
\begin{equation}
    \ln h_t=\alpha+\delta\ln h_{t-1}+\sigma_\nu\nu_t,\ t = 2,...,N
\end{equation}

At a given $t = 1,...,N$, the return from holding a financial asset equals $y_t$ and the latent log volatility $h_t$ follows the first-order auto-regressive process with $\delta$, $\alpha$ and $\sigma_\nu$ parameters, where $\alpha$ is the intercept parameter, $\delta$ is the volatility persistence and $\sigma_\nu$ is the standard deviation of the shock to the logarithm of $h_t$. 

This particular form of the model has been employed in studies by \cite{Taylor}, \cite{Hull_White}, \cite{Chesney_Scott}, \cite{Shephard}, \cite{Ghysels_Harvey_Renaul}, \cite{Jacquier}, \cite{Kim1998}. The works mentioned above extensively discuss the fundamental econometric characteristics of the model, as well as the estimation procedures for Stochastic Volatility models. 

Following the work of \cite{Jacquier}, the joint unconditional distribution including $y_t$, $h_t$, $\delta$, $\alpha$, $\sigma_\nu$ is 
\begin{equation}
    \begin{aligned}
        p(y,h,\delta,\alpha,\sigma_\nu^2)\propto & \frac{1}{\sigma_\nu^{\nu_0}\sigma_\delta\sigma_\alpha\sigma_\nu^2}\exp(-\frac{(\delta-\delta_0)^2}{2\sigma_\delta^2}-\frac{(\alpha-\alpha_0)^2}{2\sigma_\alpha^2}-\frac{s_0^2}{2\sigma_\nu^2}) \\
        & \times\prod_{t=2}^{N}\frac{1}{\sqrt{h_t}h_t\sigma_\nu}\exp(-\frac{y_t^2}{2h_t}-\frac{(\ln h_t-\delta\ln h_{t-1}-\alpha)^2}{2\sigma^2_\nu})
    \end{aligned}
\end{equation}

From this joint distribution, we can derive the posterior distributions of $\delta$, $\alpha$, $\sigma_\nu$:
\begin{equation}
    p(\sigma_\nu^2|h,\alpha,\delta)\sim IG\left(\frac{\nu_0+N-1}{2},\frac{s}{2}\right)
\end{equation}
\begin{equation}
    p(\delta|h,\alpha,\sigma_\nu^2)\sim N\left(\frac{\sigma_\nu^2\delta_0+\sigma_\delta^2 s_3-\sigma_\delta^2\alpha s_1+\sigma_\delta^2\alpha\ln h_N}{\sigma_\nu^2+\sigma_\delta^2 s_2-\sigma_\delta^2(\ln h_N)^2},\frac{\sigma_\nu^2\sigma_\delta^2}{\sigma_\nu^2+\sigma_\delta^2(s_2-(\ln h_N)^2)}\right)
\end{equation}
\begin{equation}
    p(\alpha|h,\sigma_\nu^2,\delta)\sim N\left(\frac{\sigma_\nu^2\alpha_0+\sigma_\alpha^2(1-\delta)s_1-\sigma_\alpha^2\ln h_1+\sigma_\alpha^2\delta\ln h_N}{\sigma_\nu^2+N\sigma_\alpha^2-\sigma_\alpha^2},\frac{\sigma_\nu^2\sigma_\alpha^2}{\sigma_\nu^2+N\sigma_\alpha^2-\sigma_\alpha^2}\right)
\end{equation}

where
$s=s_0+(N-1)\alpha^2+(1+\delta^2)s_2-\delta^2(\ln h_N)^2-(\ln h_1)^2-2\alpha s_1\delta + 2\alpha\ln h_1 - 2\alpha\delta\ln h_N-2\delta s_3
$, $s_1=\sum_{t=1}^N\ln h_t$, $s_2=\sum_{t=1}^N(\ln h_t)^2$, $s_3=\sum_{t=2}^N \ln h_t\ln h_{t-1}$.

The conditional posterior of $h_t$ is
\begin{equation}
    \label{eq:5}
    p(h_t|h_{t+1},h_{t-1},\delta,\alpha,\sigma_\nu^2)\propto\frac{1}{\sqrt{h_t}}\exp\left(-\frac{y_t^2}{2h_t}\right)\frac{1}{h_t}\exp\left(-\frac{(\ln h_t-\mu_t)^2}{2\sigma^2}\right)
\end{equation}

where
\begin{equation}
    \mu_t=\frac{\delta(\ln h_{t+1}+\ln h_{t-1})+\alpha(1-\delta)}{1+\delta^2}, \sigma^2=\frac{\sigma_\nu^2}{1+\delta^2}
\end{equation}

In conventional parametric SVM, the error terms, $u_t$ and $\nu_t$, are typically assumed to follow independent normal distributions, a presumption that has been increasingly scrutinized and challenged in contemporary literature. For example, \cite{Omori2007} and \cite{Jacquier2004} have proposed using a t-distribution. Similarly, \cite{Barndorff-Nielsen} introduced the Normal-Inverse Gaussian distribution as an alternative. Additionally, \cite{Mahieu_Schotman} have employed a mixture of normal distributions to better capture the underlying volatility dynamics, and \cite{Abanto-Valle2011} applied a scale mixture of Normals, incorporating varying mixing parameters. However, \cite{Data-Dependent}, \cite{One-step} argued that for GARCH models, parametric assumptions have their limitations as the parametric models do not accurately reflect the distributions of financial data. These studies collectively signify a growing recognition and adoption of more diverse and potentially more representative error distributions in modeling the volatility of financial time series.

Models with nonparametric error distributions have been demonstrated to capture characteristics of return data more effectively than their parametric counterparts, as evidenced by research from \cite{Gallant1997}, \cite{Mahieu_Schotman}, and \cite{Durham2006}. In particular, for GARCH models, \cite{One-step}, \cite{efficiency_semi_para}, \cite{Data-Dependent} proposed a nonparametric alternative that captures the salient features of the data, thereby resulting in a more efficient estimation of the volatility. These observations underscore a fundamental limitation in simple parametric models: their frequent inadequacy in accurately modeling the conditional return distribution as well as the volatility. The implication is clear: to obtain a more comprehensive representation of financial data, the adoption of nonparametric methodologies may often be necessary. This shift from traditional parametric approaches reflects a deeper understanding of the complex nature of financial markets and the need for more sophisticated analytical tools to decipher them.

In this article, for the first part, we discard the presumption of normality for $u_t$, the error term of the observation equation, instead assuming that $u_t$ adheres to a nonparametric distribution denoted by $f$. Consequently, the error terms are revised to $u_t \sim f$ while the distribution of $\nu_t$ continues to be $N(0,1)$. In other words, first, we only handle the relatively more straightforward problem of relaxing the distributional assumption of $u_t$.

For the second part, we consider the most general problem in which the error terms from the return and log-volatility equations are both assumed to be nonparametrically distributed, namely, the error term from the return, $u_t \sim f$ and the other error term $\nu_t \sim g$, where $f$ and $g$ represent unknown probability density functions. The following section will detail how these PDFs are estimated from recursively generated results and address the computational questions.

\subsection{A semiparametric formulation of the SVM}

A critical aspect of the nonparametric distribution of the error setting is what the error PDFs, namely, $f$ and $g$, should be. Here, we leverage the kernel density estimation method to approximate $f$ and $g$, which do not assume any specific functional form for the underlying distributions by \cite{Muhsal2010}. In practice, since the error distributions of $u_t$ and $\nu_t$ are unknown, we replace them with the kernel density estimation of the residuals of the Equations \ref{eq: model}, where the residuals are obtained by initially fitting the conventional parametric model proposed by \cite{Jacquier}. 

Let $y_t$ represent the asset's return. The work of \cite{Jacquier}, \cite{Jacquier2004} introduced a Bayesian method for estimating parameters and sampling volatility under the assumption of normally distributed errors. This model allows us to obtain samples of volatility, denoted as $\hat{h_0}, \hat{h_1}, ..., \hat{h_N}$, along with the posterior distributions of the parameters: $\delta$, $\alpha$, and $\sigma_\nu$. Thus, the Bayes estimates of the model parameters can be obtained as the mean of the posterior distributions of $\delta$, $\alpha$, and $\sigma_\nu$, expressed as $\hat{\delta}$, $\hat{\alpha}$, and $\hat{\sigma_\nu}$. 

\begin{figure}[H]
    \centering
    \includegraphics[width=0.45\linewidth]{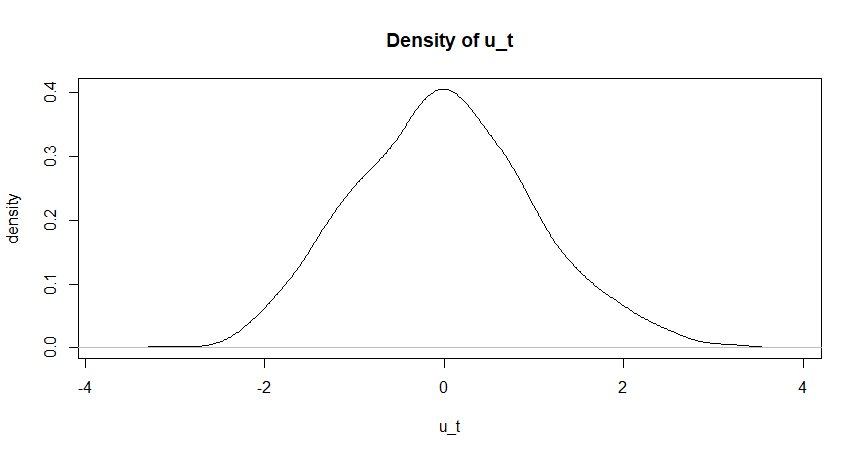}
    \includegraphics[width=0.45\linewidth]{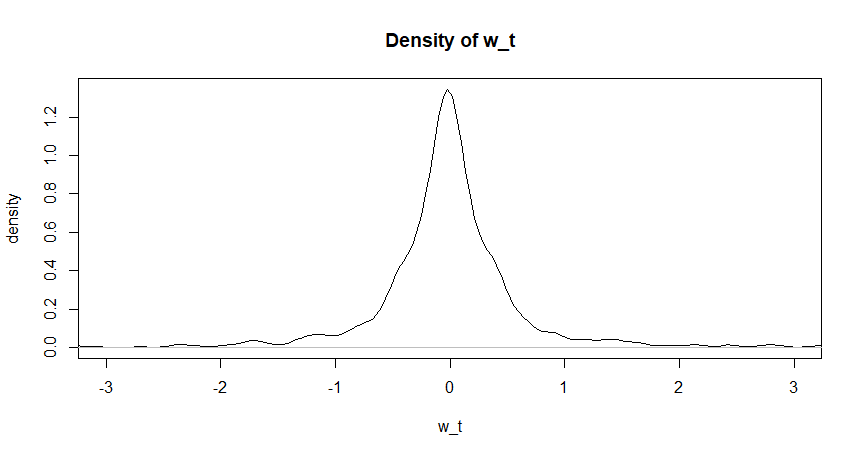}
    \caption{Estimated pdf of $u_t$ and $w_t$ based on the model residuals}
    \label{fig:u_t&w_t}
\end{figure}

Based on the sampled volatilities and the parameter estimates, we define $\hat{u_t}$ as $\hat{u_t} = \frac{y_t}{\sqrt{\hat{h_t}}}$, $t = 0,1,...,N$, and $\hat{w_t}$ as $\hat{w_t} = \frac{\ln \hat{h_t} - \hat{\mu_t}}{\hat{\sigma_\nu}}$, $t=0,1,...,N$, where $\hat{\mu_t} = \frac{\hat{\delta}\ln \hat{h_{t+1}}+\hat{\delta}\ln \hat{h_{t-1}}+(1-\hat{\delta})\hat{\alpha}}{1+\hat{\delta}^2}$, as the residuals of the return and volatility equations. Then kernel density estimates, say, $\hat{f}$ and $\hat{g}$, of the standardized residuals $\hat{u_t}$ and $\hat{w_t}$, i.e., $\frac{\hat{u_t} - \mathrm{mean}(\hat{u_t})}{\mathrm{sd}(\hat{u_t})}$ and $\frac{\hat{w_t} - \mathrm{mean}(\hat{w_t})}{\mathrm{sd}(\hat{w_t})}$, can replace the two Gaussian components of the posterior in equation \ref{eq:5}. An example of a density plot of $u_t$ and $w_t$ is shown in Figure \ref{fig:u_t&w_t}.
Therefore, if we assume the error term of the return equation, $u_t$, is nonparametrically distributed and can be evaluated by kernel density estimation, then the approximation of posterior of volatility $h_t$ is  
\begin{equation}
    \label{eq:nonpara_return_likelihood}
    p(h_t|h_{t+1},h_{t-1},\delta,\alpha,\sigma_\nu^2)\propto \frac{1}{h_t^{3/2}}\hat{f}\left(\frac{y_t}{\sqrt{h_t}}\right)\exp\left(-\frac{(\ln h_t-\mu_t)^2}{2\sigma^2}\right)
\end{equation}
Therefore, under the assumption of $u_t \sim f$ and $\nu_t \sim N(0,1)$, the likelihood is Equation \ref{eq:nonpara_return_likelihood}, we name this model as NSVM-1.
Additionally, the kernel density estimation of the two residuals provides an approximation of the new posterior of volatility $h_t$ is
\begin{equation}
    \label{eq:nonpara_vol_likelihood}
    p(h_t|h_{t+1},h_{t-1},\delta,\alpha,\sigma_\nu^2)\propto \frac{1}{h_t^{3/2}}\hat{f}\left(\frac{y_t}{\sqrt{h_t}}\right)\hat{g}\left(\frac{\ln h_t-\mu_t}{\sigma_\nu}\right)
\end{equation}
Similarly, under the assumption of $u_t \sim f$ and $\nu_t \sim g$, the likelihood is Equation \ref{eq:nonpara_vol_likelihood}, we name this model as NSVM-2.

Here $\hat{f}(x) = \frac{1}{n\times b}\sum_{i=1}^{n} N(\frac{x - x_i}{b})$, is the kernel density estimation of $\frac{\hat{u_t} - \mathrm{mean}(\hat{u_t})}{\mathrm{sd}(\hat{u_t})}$ and 
$\hat{g}(x) = \frac{1}{n\times b}\sum_{i=1}^{n} N(\frac{x - x_i}{b})$ is the kernel density estimation of $\frac{\hat{w_t} - \mathrm{mean}(\hat{w_t})}{\mathrm{sd}(\hat{w_t})}$, where $b$ is the bandwidth and $N$ is the density function of standard normal distribution. In our implementation, we use the built-in \texttt{density} function in R to obtain the kernel density estimation of Equation \ref{eq:nonpara_return_likelihood} \ref{eq:nonpara_vol_likelihood}, and the bandwidth $b$ is chosen optimally using a standard bandwidth selection method. Details about bandwidth selection is showed in Appendix.





\subsection{Algorithm}

As is often the case, the nonparametric assumption for residuals means that the posterior distribution of $h_t$ lacks a closed-form expression. Consequently, we need a numerical method to sample from this posterior distribution to estimate both volatility and model parameters. In this paper, we will utilize the Metropolis-Hastings (MH) algorithm by \cite{MH}, \cite{MH1}, \cite{MH2}, \cite{MH3} to address the challenge posed by intractable posterior distributions. This approach will also enable us to define customized proposal distributions for the specific model, enhancing the efficiency of the sampling process. Moreover, the MH algorithm demonstrates robust theoretical convergence by \cite{ConvergenceRates_MCMC}, \cite{LB_ConvergenceRates}, assuring reliable outcomes under mild assumptions, making it a strong candidate for posterior sampling. 

Algorithm \ref{alg:sketch_main} presents a sketch of the primary algorithm. See Appendix: Algorithm \ref{Alg:sampler_main} for a comprehensive discussion of the main sampler function.

\begin{algorithm}[H]
    \caption{Sketch main algorithm} 
    \begin{algorithmic}
        \Require $Input: (y,v_0,s_0,\delta_0,\sigma_\delta^2,\alpha_0,\sigma_\alpha^2,T,b)$
        \State Initialize parameters $\delta$, $\alpha$, $\sigma_\nu$ and $h$
        \For{i = 1,...,T+b}
            \For{t = 1,...,N-1}
                \State Draw $h_t$ from its posterior distribution
            \EndFor
            \State Let $h_0$ be the value such that $ln\ h_1 = \alpha + \delta ln\ h_{0} + \sigma_\nu \nu_1$
            \State Let $ln\ h_N = \alpha + \delta ln\ h_{N-1} + \sigma_\nu \nu_N$
            \State Draw $\delta$ from its posterior distribution
            \State Draw $\alpha$ from its posterior distribution
            \State Draw $\sigma_\nu$ from its posterior distribution
        \EndFor
    \end{algorithmic}
    \label{alg:sketch_main}
\end{algorithm}

\subsubsection{Semiparametric SVM algorithm with estimated pdf \texorpdfstring{$f$}{f}}

Nonparametric Bayesian methods offer considerable flexibility by allowing model complexity to grow with the data, adapting to the underlying structure without predefined assumptions about the number of parameters. This adaptability is particularly advantageous when the true structure of the data is unknown, as it enables the model to capture intricate patterns that parametric models may overlook.

If we relax the parametric normality assumption on $u_t$ and replace it with the kernel estimate of the pdf $f$ while maintaining the assumption that $\nu_t$ follows a standard normal distribution, this alteration will not influence the conditional posterior distributions of the parameters $\delta$, $\alpha$, and $\sigma_\nu$, based on the parametric prior proposed by \cite{Jacquier}, i.e., $\alpha\sim N(\alpha_0,\sigma_\alpha^2)$, $\delta\sim N(\delta_0,\sigma_\delta^2)$,
$\sigma_\nu^2\sim IG(\frac{\nu_0}{2},\frac{s_0}{2})$. Given the parametric form of these priors, sampling from these distributions remains straightforward. 

The sampling approach introduced by \cite{Jacquier} provides an efficient mechanism for enhancing the updating step in the Metropolis-Hastings (MH) algorithm by \cite{MH}, \cite{MH2}. This improvement hinges on an informed approximation of a specific distribution achieved through an inverse gamma distribution. The motivation for this selection is to match the first and second moments of the proposed distribution with those of the log-normal component in the posterior distribution. This alignment is established by calibrating the parameters of the inverse gamma distribution as follows:

The proposal distribution $q(h_t)$ is defined by the equation:
\begin{equation}
    q(h_t)=\frac{\lambda^\phi}{\Gamma(\phi)}h^{-(\phi+1)}e^{-\frac{\lambda}{h_t}}
\end{equation}

where
$
\lambda=\frac{1-2e^{\sigma^2}}{1-e^{\sigma^2}}+\frac{1}{2}
$ and 
$
\phi=(\lambda-1)e^{\mu_t+\frac{\sigma^2}{2}}+\frac{y_t^2}{2}
$ are the parameters of inverse gamma distribution.


The sketch algorithm for sampling the volatility is shown in Algorithm \ref{alg:sketch_sampler_h}.

\begin{algorithm}[H]
    \caption{Sketch algorithm of sampler for $h$}  
    \begin{algorithmic}
        \Require $Input: (y,\alpha,\delta,\sigma_\nu^2,ln\_h_{i-1},h_i,ln\_h_{i+1})$
        \State Calculate $\mu$, $\sigma^2$, $\lambda$, $\phi$ by using input values
        \While{not accepted}
            \State Draw the proposed $h_{new}\sim IG\left( \lambda,\phi \right)$
            \State Accept $h_{new}$ with probability $Min\left(1,\frac{p(h_{new})}{c q(h_{new})} \right)$, where $q$ is the density of inverse gamma distribution.
        \EndWhile
        \If{$p(h_{new})<c q(h_{new})$}
            \State Return $h_{new}$
        \EndIf
        \State Accept $h_{new}$ with probability $Min\left(1,\frac{p(h_{new})/q(h_{new})}{p(h_{old})/q(h_{old})} \right)$, where $h_{old}$ is previous sample of $h_t$
        \If{accepted}
            \State Return $h_{new}$
        \Else 
            \State Return $h_{old}$
        \EndIf
    \end{algorithmic}
    \label{alg:sketch_sampler_h}
\end{algorithm}

Here the constant $c$ is defined as:
\begin{equation}
    \label{eq:c_star}
    c=c^\star\left(\frac{p(h_m)}{q(h_m)}\right)
\end{equation}
 
where $h_m$ is the mode of $q$ and $c^\star$ here is a tuning parameter. This constant serves as a scaling factor, ensuring that the proposed distribution $q(h_t)$ closely mirrors the target distribution $p(h)$ in terms of its mode. In our implementation in the simulation, we choose $c^\star = 1.2$, where by comparison of results in terms of bias, details about tuning $c^\star$ are shown in the Appendix. This nuanced calibration enhances the accuracy of the approximation and contributes to the overall efficiency and efficacy of the MH algorithm in navigating the complex landscape of the distribution under study. The detailed algorithm for sampling the volatility is included in Appendix: Algorithm \ref{alg:sampler_h}.

\subsubsection{Semiparametric SVM with independent unknown error distributions}

In a real-world setting, the distribution of the error term in the return equation and the error term in the volatility equation are both unknown. Therefore, we utilize the same methodology to estimate them from data. By allowing the model’s complexity to evolve with real-world data, nonparametric Bayesian models provide more accurate uncertainty quantification.

If we assume the distributions of both error terms $u_t$ and $\nu_t$ are unknown, then nonparametric estimates of the joint distribution of fitted residuals $\hat{u_t}$ and $\hat{\nu_t}$ are needed. Consequently, the posterior distributions of $\delta$, $\alpha$, and $\sigma_\nu$ no longer have a parametric form. Therefore, numerical methods like MCMC must be utilized to sample from their nonparametric posterior distributions. Here, we use parameter $\delta$ as an example; similar sampling procedures are deployed on other parameters.

The sketch of the sampler algorithm for nonparametric $\delta$ is shown in Algorithm \ref{alg:sketch_sampler_delta}.

\begin{algorithm}[H]
    \caption{Sketch algorithm of sampler for $\delta$}
    \begin{algorithmic}
        \Require $Input: (\delta_0,\sigma_\delta^2,\alpha,\sigma_\nu^2,h,\delta_{left})$
        \State Let proposal distribution $q$ be a normal distribution with the same mean and variance in the parametric setting
        \State Let $p$ be the posterior distribution in the parametric setting
        \While{not accepted}
            \State Draw proposed $\delta_{new}$ from proposal distribution
            \State Accept $\delta_{new}$ with probability $Min\left(1,\frac{p(\delta)}{q(\delta)}\right)$
        \EndWhile
        \If{$p(\delta)<q(\delta)$}
            \State \Return $\delta_{new}$
        \EndIf
        \State Accept $\delta_{new}$ with probability $Min\left(1, \frac{p(\delta_{new})/q(\delta_{new})}{p(\delta_{left})/q(\delta_{left})} \right)$, where $\delta_{left}$ is the previous sample of $\delta$
        \If{accepted}
            \State \Return $\delta_{new}$
        \Else
            \State \Return $\delta_{left}$
        \EndIf
    \end{algorithmic}
    \label{alg:sketch_sampler_delta}
\end{algorithm}

The detailed algorithms for sampling posterior of parameters are included in Appendix: Algorithm \ref{alg:sampler_sigma}, \ref{alg:sampler_delta} and \ref{alg:sampler_alpha}.

As mentioned in \ref{eq:nonpara_vol_likelihood}, the likelihood of volatility $h_t$ is also based on nonparametric estimates of the pdfs of the error terms $u_t$ and $\nu_t$. Thus, we utilize the same algorithm \ref{alg:sketch_sampler_h} but revise the posterior distribution of the volatility sequence $h_t$ from the equation \ref{eq:nonpara_return_likelihood} to equation \ref{eq:nonpara_vol_likelihood}.

\section{Simulations and Application}

This section examines the performance of the proposed method through simulations and an analysis of the S\&P 500 data.

\subsection{Simulation analysis}

This section presents the results of applying our model to simulated data. In real-world scenarios, where the true values of volatility and the three parameters remain unknown, evaluating the performance differences between parametric and nonparametric models is quite challenging. In contrast, a generated simulation can offer a known ground truth for the estimated parameter values and realized volatility, enabling a more detailed comparison of model performance.

\subsection{Simulation for Gaussian model}

The objective here is to assess the performance of the proposed semiparametric method when the data-generating process involves normally distributed errors. In particular, we set $\alpha = -0.15$, $\delta = 0.985$, $\sigma = 0.15$. 

Initialize $h_0\sim \mathrm{N}\left(\frac{\alpha}{1-\delta}=-10,\frac{\sigma}{\sqrt{1-\delta^2}}=0.87\right)$, $y_0 \sim \sqrt{h_0} * \mathrm{N}\left(0,1\right)$. For $i = 1,2,...,500$, $y_i =\sqrt{h_i}u_i$, $ln\left(h_{i}\right) = \alpha + \delta ln\left(h_{i-1}\right) + \sigma_\nu \nu_i$. Here $u_i \sim \mathrm{N}\left(0,1\right)$ and $\nu_i \sim \mathrm{N}\left(0,1\right)$. We run each MCMC for 10,000 iterations with the first 5,000 iterations designated as burn-in. Only the samples obtained after the 5,000-th iteration were retained for subsequent analysis.

\begin{figure}[H]
    \centering
    \includegraphics[width=\linewidth]{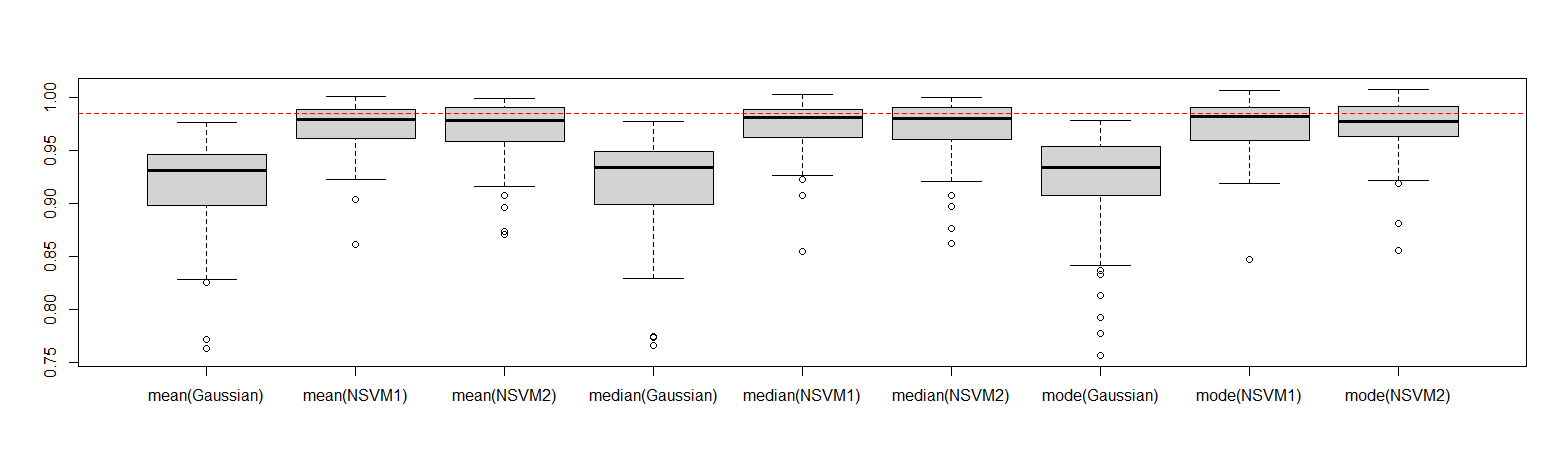}
    \caption{Estimation of delta under Gaussian error model simulation. The dotted line is the true value (0.985)}
    \label{fig:100run_delta_gaussian}
\end{figure}

\begin{figure}[H]
    \centering
    \includegraphics[width=\linewidth]{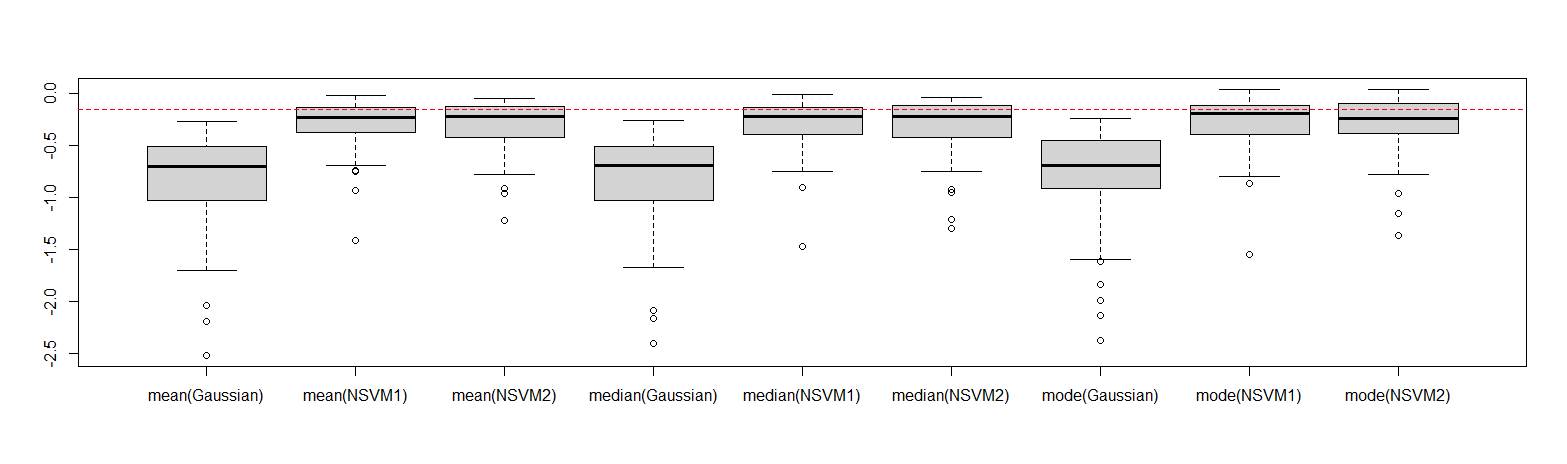}
    \caption{Estimation of delta under Gaussian error model simulation. The dotted line is the true value (-0.15)}
    \label{fig:100run_alpha_gaussian}
\end{figure}

\begin{figure}[H]
    \centering
    \includegraphics[width=\linewidth]{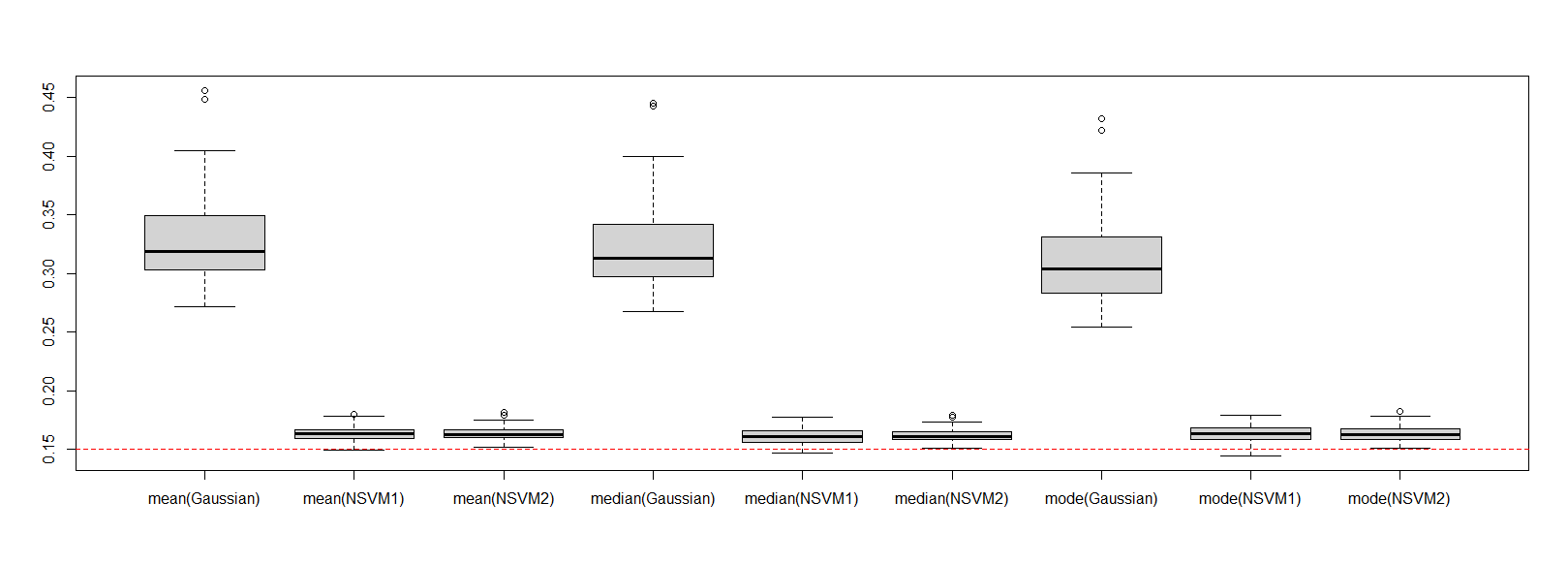}
    \caption{Estimation of sigma under Gaussian error model simulation. The dotted is the true value (0.15)}
    \label{fig:100run_sigma_gaussian}
\end{figure}

To assess the precision and accuracy in parameter estimation, we repeated this process 100 times, each time we generated new $y_1, ..., y_{500}$ with the same parameter values and recorded the result: the mean, median, and mode of the posterior distributions, $p(\alpha|h,\delta,\sigma_\nu^2)$, $p(\sigma_\nu^2|h,\alpha,\delta)$, and $p(\delta|h,\alpha,\sigma_\nu^2)$. Figures \ref{fig:100run_delta_gaussian}, \ref{fig:100run_alpha_gaussian} and \ref{fig:100run_sigma_gaussian} present the results for the Gaussian, NSVM-1, and NSVM-2 models, respectively. The red dashed lines indicate the true values used in generating the data.

As shown in the box plots, the results from the NSVM-1 and NSVM-2 models exhibit a minor deviation from the red dashed line, representing the actual value than those of the Gaussian model, indicating lower bias. Furthermore, the narrower range of the box plots for NSVM-1 and NSVM-2 suggests a substantially lower variance of these estimates. Additionally, the plots reveal no significant differences among the distributions' mean, median, or mode. To quantitatively assess performance, we define the mean square error (MSE) of them can be defined as $\frac{1}{100}\left(\Sigma \bar{\delta_i} - \delta \right)^2$, $\frac{1}{100}\left( \Sigma \bar{\alpha_i} - \alpha \right)^2$ and $\frac{1}{100}\left( \Sigma \bar{\sigma_{\nu i}} - \sigma_\nu \right)^2$. For the Gaussian model, the MSE of mean, median, and mode for $\delta$ are all approximately 0.004, whereas the MSEs for NSVM-1 and NSVM-2 are 0.000156 and 0.000213, respectively. For $\alpha$, the MSE for the Gaussian model is 0.4526, while for NSVM-1 and NSVM-2, it is 0.0297 and 0.0363, respectively. For $\sigma_\nu$, the Gaussian model yields an MSE of 0.0314, whereas the NSVM-1 and NSVM-2 models yield significantly lower MSE values of 0.000187 and 0.000183, respectively.

The results show that the proposed semiparametric methods outperform the traditional parametric SVM, even in the ideal setup where the data-generating process is Gaussian and the fitted model is correctly specified. This demonstrates that a parametric model cannot account for the anomalies in the data, while the semiparametric approaches offer a superior alternative by providing better accounting for the features of the data.

\subsection{Simulations for heavy-tailed error distributions}

Alongside simulations using normally distributed error settings, we assess the model's performance with heavy-tailed error distributions, including the Student's t distribution and the Generalized Error Distribution (GED). Analyzing heavy-tailed error distributions is essential since real-world data frequently display leptokurtic traits that diverge from the Gaussian norm. Models demonstrating strong performance under these distributions are more applicable in fields like finance and economics, where such distributions are prevalent, thereby enhancing the robustness and versatility of volatility estimation techniques. In the following discussion, we will maintain the same parameter values as in the prior section; the sole variation is in the error term distributions, $u_i$ and $
u_i$.

\subsubsection{Student's t distributed error simulation}

\begin{figure}[H]
    \centering
    \includegraphics[width=\linewidth]{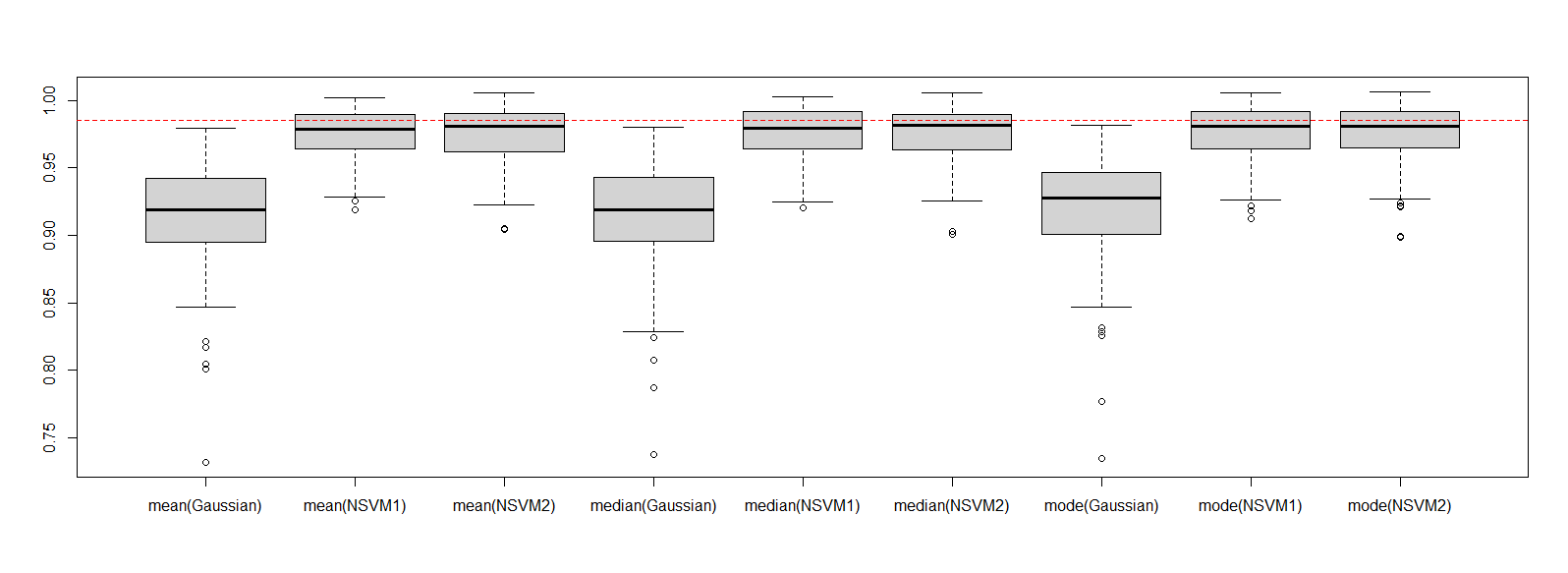}
    \caption{Estimation of delta under Student's t-distribution error model simulation. The dotted line is the true value (0.985)}
    \label{fig:100run_delta_t}
\end{figure}

\begin{figure}[H]
    \centering
    \includegraphics[width=\linewidth]{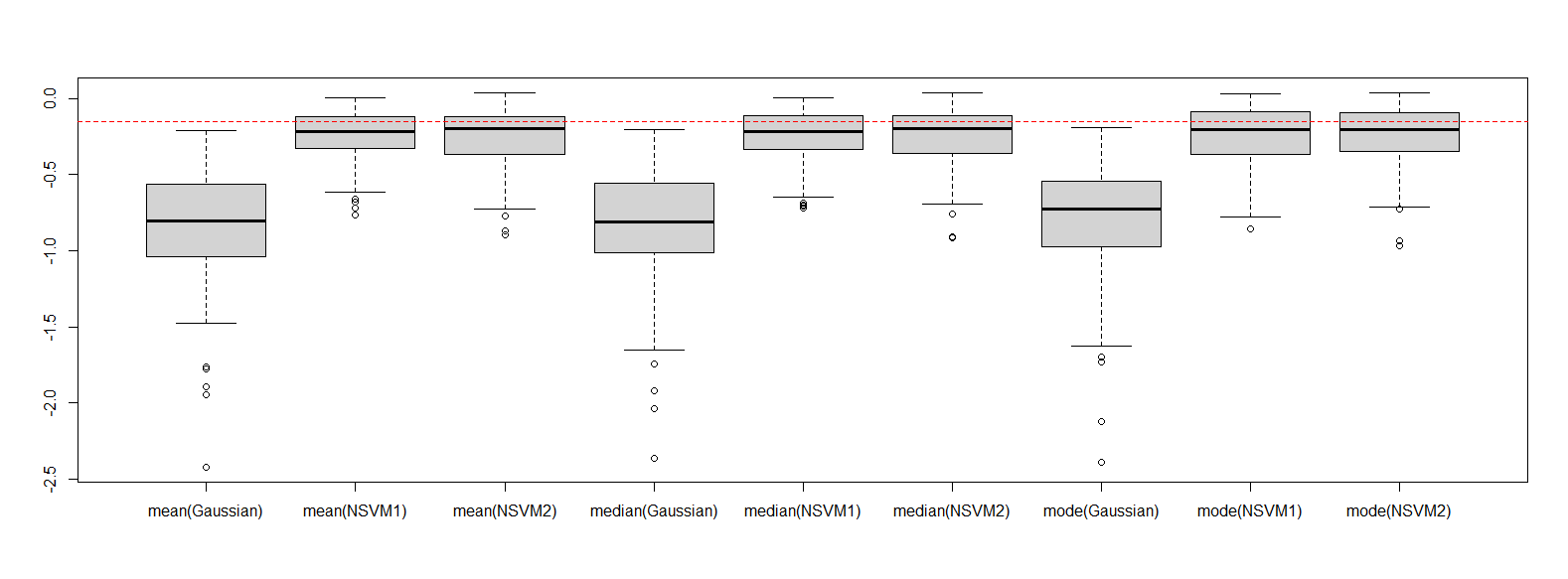}
    \caption{Estimation of alpha under Student's t-distribution error model simulation. The dotted line is the true value (-0.15)}
    \label{fig:100run_alpha_t}
\end{figure}

\begin{figure}[H]
    \centering
    \includegraphics[width=\linewidth]{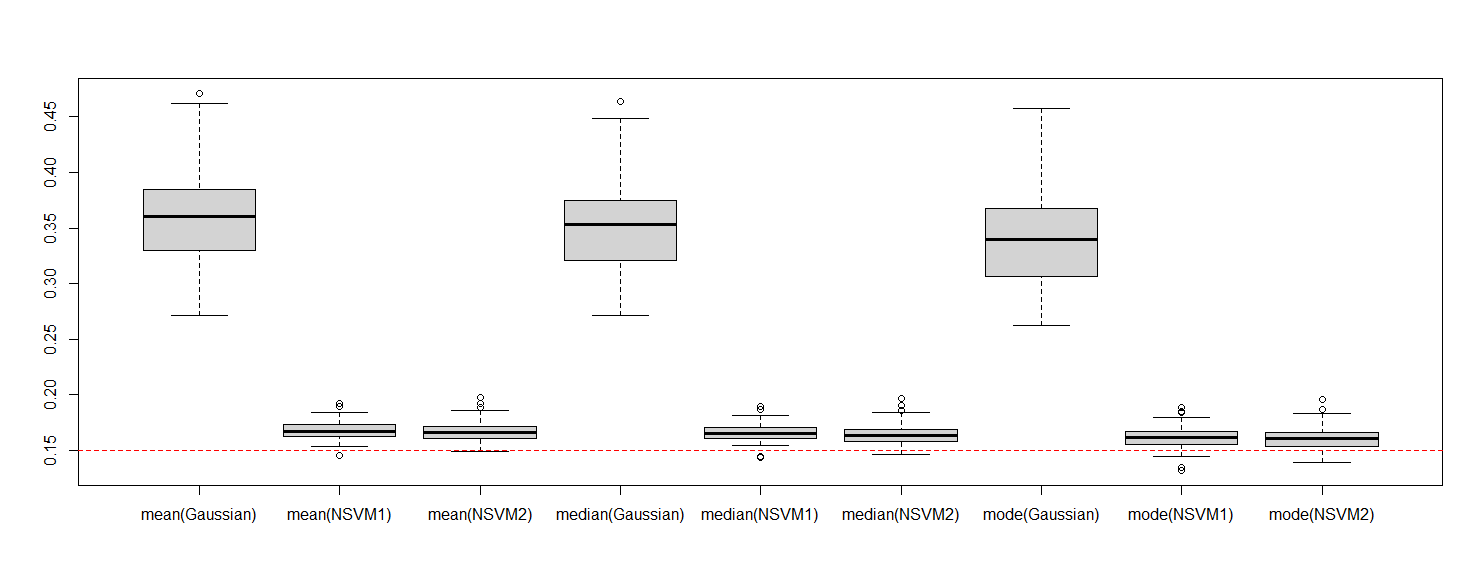}
    \caption{Estimation of sigma under Student's t-distribution error model simulation. The dotted line is the true value (0.15)}
    \label{fig:100run_sigma_t}
\end{figure}

The Student's t distribution, especially with low degrees of freedom, is recognized for its heavy tails, which increase the chances of extreme values compared to the normal distribution. The heaviness of the tails is influenced by the degrees of freedom; fewer degrees lead to heavier tails, allowing the distribution to approximate normality as the degrees of freedom rise 
\cite{Studentt_raanju}. In our simulation, we set $u_i$ and $
u_i$ to follow a Student's t distribution with 10 degrees of freedom. Figures \ref{fig:100run_delta_t}, \ref{fig:100run_alpha_t}, and \ref{fig:100run_sigma_t} show the results of the Student's t distributed error simulation for the Gaussian, NSVM-1, and NSVM-2 models, respectively, with the red dashed lines indicating the true values used to generate the data.

In line with the findings from the Gaussian simulation, the NSVM-1 and NSVM-2 models exhibit a smaller deviation from the red dashed line, which denotes the true parameter value, compared to the Gaussian model, thus indicating a reduction in bias. Furthermore, the box plots for NSVM-1 and NSVM-2 present a narrower range, signifying considerably less variance in these estimates. The plots show no significant differences regardless of whether the mean, median, or mode is used to assess central tendency. For the Gaussian model, the mean squared error (MSE) for the mean, median, and mode of $\delta$ is about 0.005, while the MSEs for NSVM-1 and NSVM-2 are 0.0001008 and 0.0001076, respectively. In terms of $\alpha$, the MSE for the Gaussian model stands at 0.5431, in contrast to 0.024 for NSVM-1 and 0.025 for NSVM-2. For $\sigma_\nu$, the Gaussian model yields an MSE of 0.04408, while NSVM-1 and NSVM-2 deliver significantly lower MSE values of 0.00031 and 0.00028, respectively.

\subsubsection{Simulations based on GED error}

\begin{figure}[H]
    \centering
    \includegraphics[width=\linewidth]{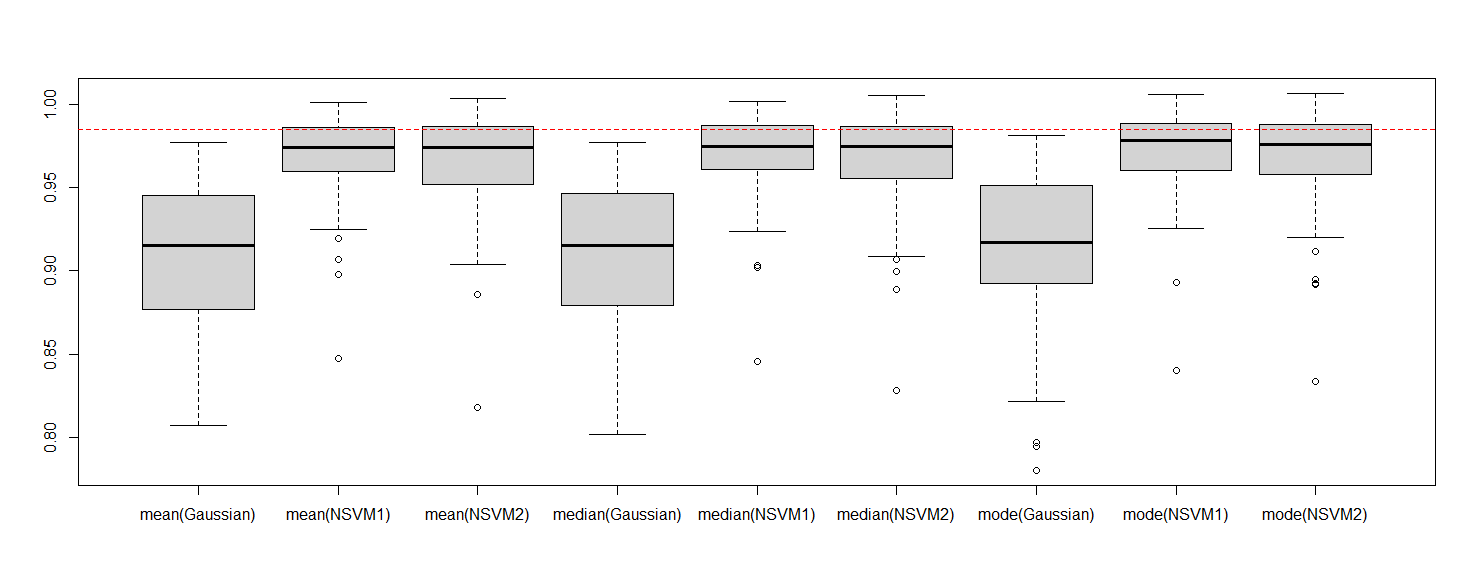}
    \caption{Estimation of delta under GED error model simulation. The dotted line is the true value (0.985)}
    \label{fig:100run_delta_ged}
\end{figure}

\begin{figure}[H]
    \centering
    \includegraphics[width=\linewidth]{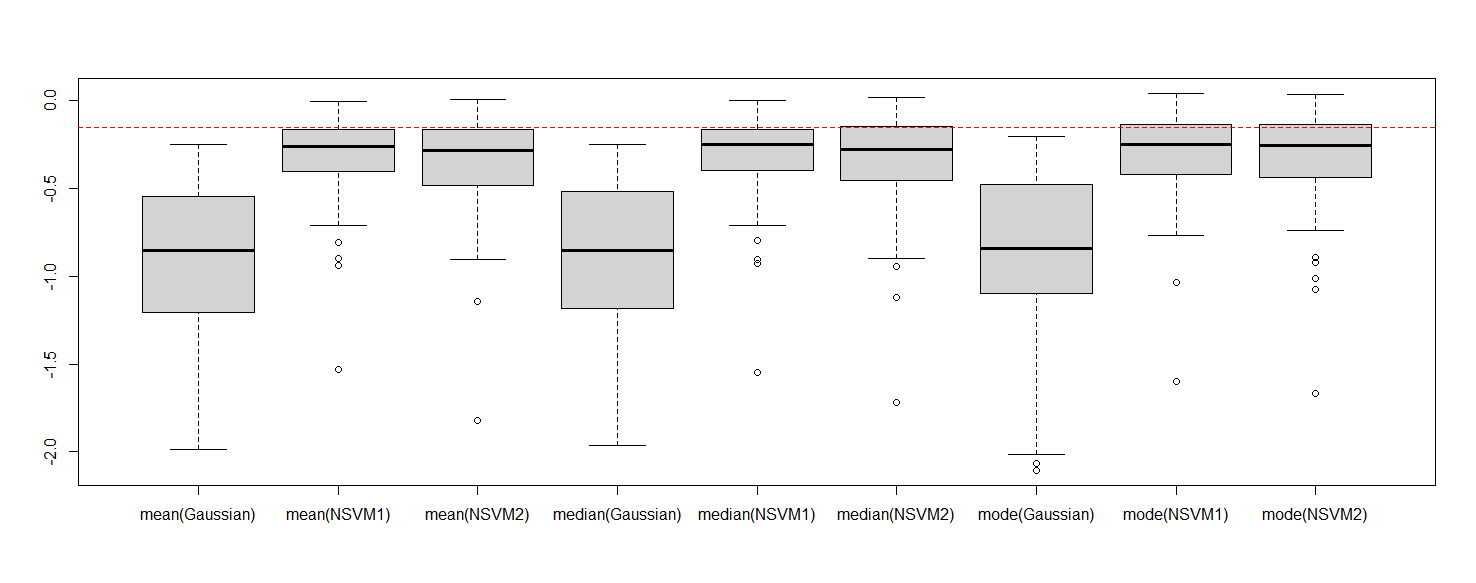}
    \caption{Estimation of alpha under GED error model simulation. The dotted line is the true value (-0.15)}
    \label{fig:100run_alpha_ged}
\end{figure}

\begin{figure}[H]
    \centering
    \includegraphics[width=\linewidth]{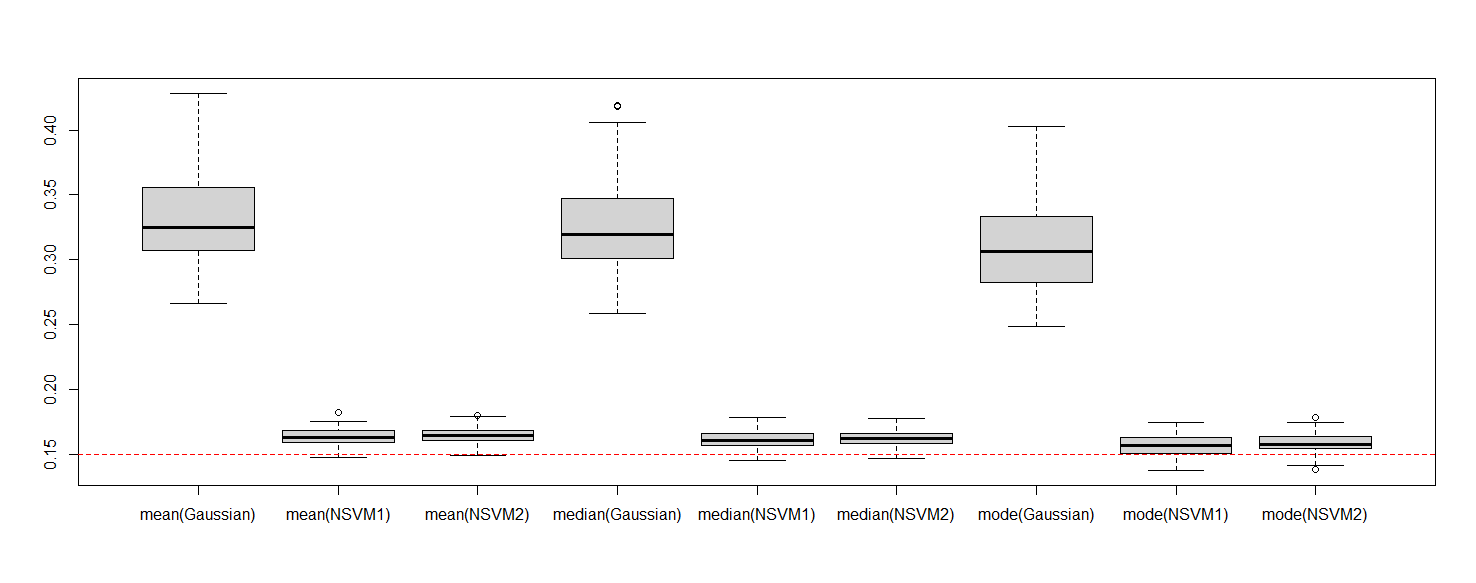}
    \caption{Estimation of sigma under GED error model simulation. The dotted line is the true value (0.15)}
    \label{fig:100run_sigma_ged}
\end{figure}

The Generalized Error Distribution (GED) is a continuous probability distribution defined by its adaptable tail thickness, offering enhanced flexibility in modeling heavy-tailed data 
\cite{GED_leopoldo}. In contrast to the normal distribution, the GED features a shape parameter that regulates the kurtosis. As this parameter diminishes, the tails grow heavier, making it appropriate for datasets with frequent large deviations from the mean. This distribution is frequently utilized in financial modeling and various domains where extreme values and outliers are common. In the simulation setup, we define $u_i$ and $\nu_i$ to adhere to a GED distribution characterized by a mean of 0 and a variance of 1. The results from the GED-distributed error simulation for the Gaussian, NSVM-1, and NSVM-2 models are shown in Figures \ref{fig:100run_delta_ged}, \ref{fig:100run_alpha_ged}, and \ref{fig:100run_sigma_ged}, respectively. As previously mentioned, the red dashed lines indicate the true values utilized for data generation.

The plot reveals steady trends, indicating that the NSVM-1 and NSVM-2 models perform better than the Gaussian model in terms of bias and variance. This is clear from the mean squared error (MSE) comparisons among all three models. For the parameter $\delta$, the Gaussian model reports MSE values around 0.0058 for the mean, median, and mode, reflecting a greater bias and variability in these estimates. In comparison, the NSVM-1 and NSVM-2 models demonstrate significantly lower MSE values of 0.00023 and 0.00035, respectively, highlighting their enhanced precision in parameter estimation.

For $\alpha$, the Gaussian model shows a notably higher MSE of 0.6663, signifying substantial estimation errors. In contrast, the NSVM-1 and NSVM-2 models achieve significantly lower MSE values of 0.0458 and 0.0616, respectively, suggesting these models are better at accurately capturing the true parameter values while reducing errors.

For $\sigma_\nu$, performance differences stand out clearly. The Gaussian model yields a mean squared error (MSE) of 0.03348, significantly surpassing the lower MSE values of NSVM-1 and NSVM-2 at 0.000175 and 0.000211, respectively. This evidence highlights the superior ability of NSVM-1 and NSVM-2 to deliver more accurate and reliable estimates, particularly in capturing nuanced variations while ensuring robustness in parameter estimation. This overall enhancement in performance illustrates the benefits of the NSVM frameworks compared to traditional Gaussian modeling methods.

\subsection{Volatility Estimation}

\begin{figure}[H]
    \centering
    \includegraphics[width=\linewidth]{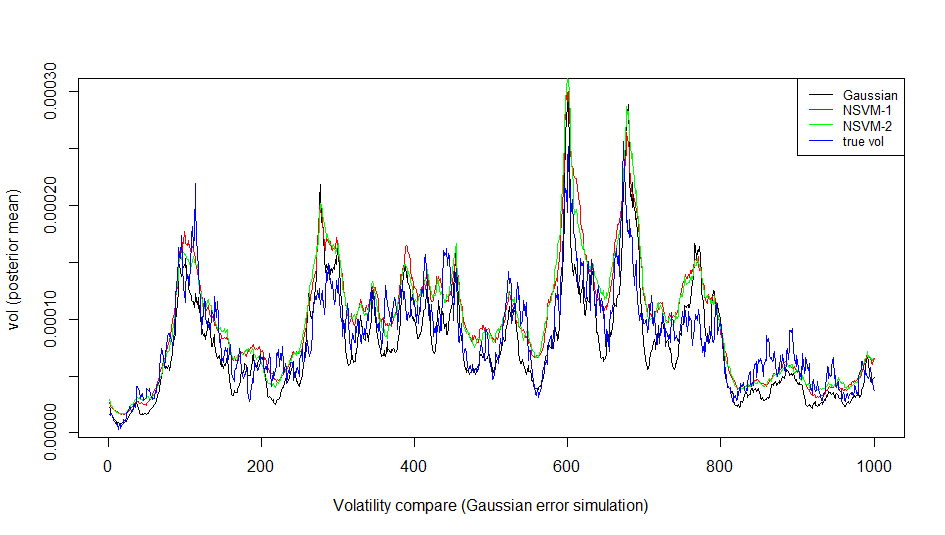}
    \caption{Comparison of volatility for Gaussian error simulation)}
    \label{fig:vol_compare_Gaussian}
\end{figure}

To assess the model's effectiveness in estimating volatility, we create a dataset using the same parameters as outlined earlier: $\alpha= -0.10$, $\delta = 0.985$, $\sigma = 0.15$. The algorithm is run with the identical data input for 100 repetitions. Each run involves executing the MCMC algorithm for 10,000 iterations, with 5,000 iterations considered as burn-in, hence keeping only the latter half of the samples. We collect the final 5,000 sampled volatility values from the posterior distribution. After the simulation concludes, we calculate the mean of these recorded volatility samples to provide the estimated volatility. This method utilizes Gaussian error simulation, and the estimated volatility from the Gaussian model, as well as NSVM-1 and NSVM-2 models, is illustrated in Figure \ref{fig:vol_compare_Gaussian}. 

\begin{figure}[H]
    \centering
    \includegraphics[width=\linewidth]{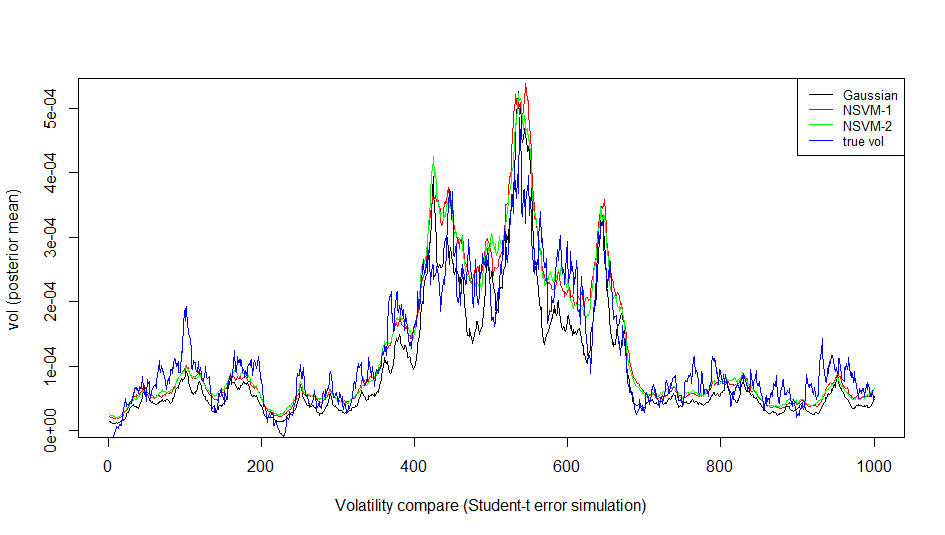}
    \caption{Comparison of volatility for Student's t-distribution error simulation)}
    \label{fig:vol_compare_t}
\end{figure}

For Student-t error simulation, the comparison is shown in Figure \ref{fig:vol_compare_t}.

\begin{figure}[H]
    \centering
    \includegraphics[width=\linewidth]{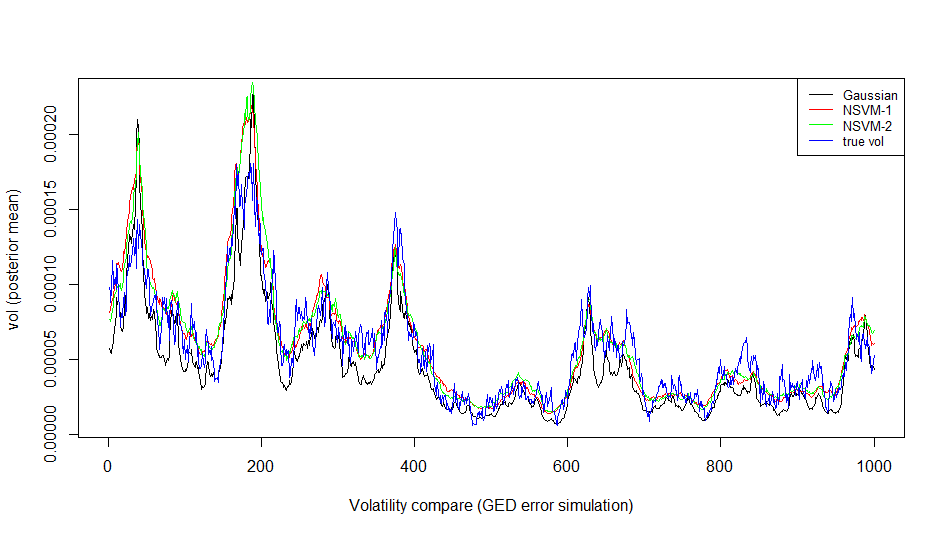}
    \caption{Comparison of volatility for GED error simulation)}
    \label{fig:vol_compare_ged}
\end{figure}

For GED error simulation, the comparison is shown in Figure \ref{fig:vol_compare_ged}.

Figures \ref{fig:vol_compare_Gaussian}, \ref{fig:vol_compare_t}, and \ref{fig:vol_compare_ged} show that the NSVM-1 and NSVM-2 models outperform the traditional Gaussian model in estimating volatility. For example, in Figure \ref{fig:vol_compare_t}, at around $t=500$ and $t=600$, the Gaussian model's volatility estimates are significantly lower than the actual values, while NSVM-1 and NSVM-2 provide estimates that closely match the true volatility. A consistent trend can be seen in Figure \ref{fig:vol_compare_ged}, especially near $t=220$ and $t=380$, where NSVM-1 and NSVM-2 accurately capture the true volatility, unlike the Gaussian model, which falls short.

This discrepancy arises from the Gaussian model's assumption about the error term, which does not account for the heavy-tailed distribution utilized in the simulation. On the other hand, the NSVM-1 and NSVM-2 models are more appropriate for situations with heavy-tailed errors, showcasing their resilience and flexibility. These findings underscore the benefits of employing NSVM frameworks, whether dealing with heavy-tailed error distributions or when the error distribution is Gaussian and the parametric model is accurately defined.

Additionally, we use the following metrics to assess and compare volatility estimation performance. Let the actual volatility from the simulation be represented as $h_1,...,h_N$, and the estimated volatility as $\bar{h_1},...,\bar{h_N}$. Besides the mean, we also take into account the median and mode of the final sampled volatilities for comparison. The three metrics used to assess volatility estimation performance are: square root of mean square error (srMSE), $\sqrt{\frac{\Sigma (h_i - \bar{h_i})^2}{N}}$, mean absolute error (MAE), $\frac{\Sigma |h_i - \bar{h_i}|}{N}$, and mean absolute percent error (MAPE), $\frac{1}{N} \Sigma 
 \frac{|h_i - \bar{h_i}|}{h_i}\times 100\%$. The table \ref{tab:vol_compare_error_gaussian} presents the results of Gaussian error simulation. Table \ref{tab:vol_compare_error_t} displays the outcomes of the Student's t error simulation, while Table \ref{tab:vol_compare_error_ged} presents the results of the GED error simulation.

\begin{table}[H]
    \centering
    \resizebox{\textwidth}{!}{
    \begin{tabular}{|c|c|c|c|c|c|c|c|c|c|c|}
        \hline
        Model  & srMSE Mean & srMSE Median & srMSE Mode & MAE Mean & MAE Median & MAE Mode & MAPE Mean & MAPE Median & MAPE Mode \\
        \hline
        Gaussian &  0.008278  & 0.008287 & 0.008296 & 0.007433 & 0.007341 & 0.007347 & \textbf{0.09932} & \textbf{0.09908} & 0.09922 \\
        NSVM-1  &  0.008252  & 0.008260 & 0.008271 & 0.007313 & 0.007325 & 0.007327 & 0.09945 & 0.09927 & \textbf{0.09912} \\
        NSVM-2  &  \textbf{0.008250}  & \textbf{0.008259} & \textbf{0.008269} & \textbf{0.007213} & \textbf{0.007318} & \textbf{0.007314} & 0.09959 & 0.09939 & 0.09944 \\
        \hline
    \end{tabular}}
    \caption{Volatility estimation error comparison (Gaussian error simulation)}
    \label{tab:vol_compare_error_gaussian}
\end{table}

\begin{table}[H]
    \centering
    \resizebox{\textwidth}{!}{
    \begin{tabular}{|c|c|c|c|c|c|c|c|c|c|c|}
        \hline
        Model  & srMSE Mean & srMSE Median & srMSE Mode & MAE Mean & MAE Median & MAE Mode & MAPE Mean & MAPE Median & MAPE Mode \\
        \hline
        Gaussian &  0.006675  & 0.006682 & 0.006691 & 0.006019 & 0.006025 & 0.006031 & 0.09959 & 0.09968 & 0.09977 \\
        NSVM-1  &  \textbf{0.006648}  & 0.006656 & 0.006677 & \textbf{0.005997} & 0.006005 & 0.006011 & \textbf{0.09948} & \textbf{0.09957} & \textbf{0.09965} \\
        NSVM-2  &  0.006651  & \textbf{0.006651} & \textbf{0.006668} & 0.005998 & \textbf{0.006003} & \textbf{0.006009} & 0.09949 & 0.09959 & 0.09966 \\
        \hline
    \end{tabular}}
    \caption{Volatility estimation error comparison (Student-t error simulation)}
    \label{tab:vol_compare_error_t}
\end{table}

\begin{table}[H]
    \centering
    \resizebox{\textwidth}{!}{
    \begin{tabular}{|c|c|c|c|c|c|c|c|c|c|c|}
        \hline
        Model  & srMSE Mean & srMSE Median & srMSE Mode & MAE Mean & MAE Median & MAE Mode & MAPE Mean & MAPE Median & MAPE Mode \\
        \hline
        Gaussian &  0.006408  & 0.006411 & 0.006416 & 0.005947 & 0.005951 & 0.005955 & 0.09967 & 0.09973 & 0.09979 \\
        NSVM-1  &  0.006395  & 0.006398 & 0.006404 & 0.005935 & \textbf{0.005938} & 0.005943 & \textbf{0.09964} & \textbf{0.09969} & \textbf{0.09973} \\
        NSVM-2  &  \textbf{0.006394}  & \textbf{0.006396} & \textbf{0.006403} & \textbf{0.005934} & 0.005939 & \textbf{0.005939} & 0.09966 & 0.09971 & 0.09977 \\
        \hline
    \end{tabular}}
    \caption{Volatility estimation error comparison (GED error simulation)}
    \label{tab:vol_compare_error_ged}
\end{table}

The tables 
\ref{tab:vol_compare_error_gaussian}, 
\ref{tab:vol_compare_error_t}, and 
\ref{tab:vol_compare_error_ged} illustrate that the volatilities estimated by the NSVM-1 and NSVM-2 models show lower square root mean squared error (srMSE), mean absolute error (MAE) and mean absolute percentage error (MAPE). This trend holds true except for the MAPE derived from the mean and median of the final sampled volatilities for the Gaussian error simulation. These reduced levels of bias offer assurance that NSVM-1 and NSVM-2 offer enhanced accuracy in estimating volatility.

\subsection{Empirical application}

This section presents the results from using our model on daily stock return data. The S$\&$P 500 (Standard and Poor's 500) is a stock market index that measures the performance of 500 of the largest publicly traded companies in the U.S. It is considered one of the best indicators of the U.S. stock market's overall performance and the economy at large. Here, we analyze the daily closing price of the S$\&$P 500 from February 1, 2021, to February 1, 2024. The simulation section demonstrates that the model also provides sampled volatility and estimated parameter values. We employed the Gaussian model, NSVM-1, and NSVM-2, using the log return of the closing price (\(\log \left( Price_{t}/Price_{t-1} \right)\)) as input. Each of the three models was executed 100 times, running MCMC for 10,000 iterations with 5,000 burn-ins; this means we only retained samples after the 5,000th iteration. For volatility estimation, we captured the last sample from the 5,000 sampled values of the volatility time series from the posterior distribution.

\begin{figure}[H]
    \centering
    \includegraphics[width=\linewidth]{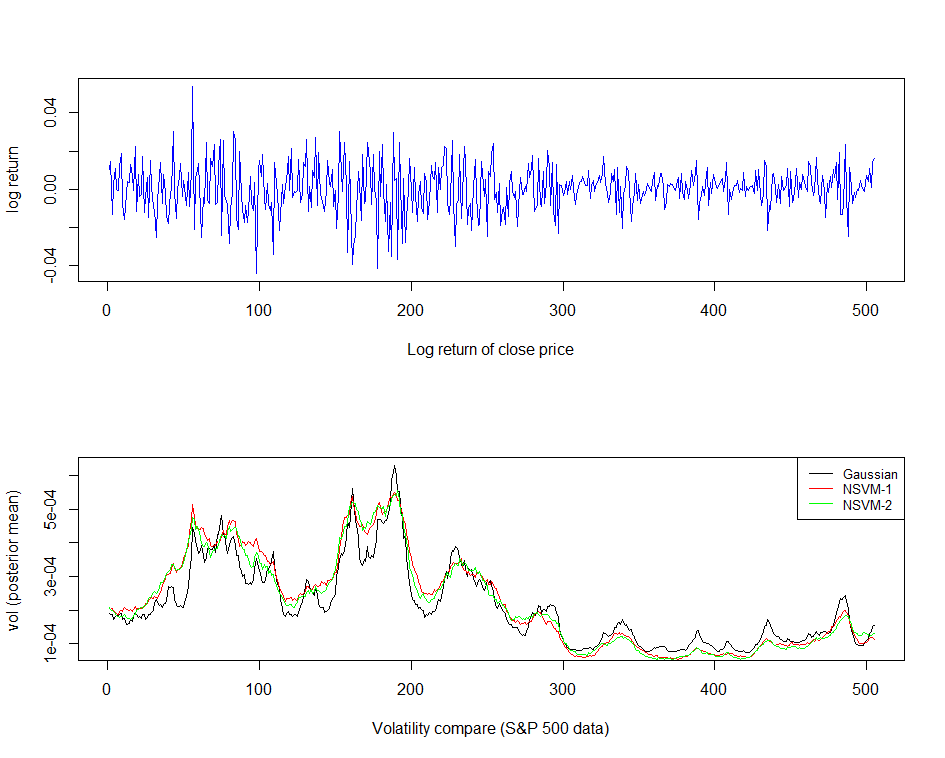}
    \caption{Comparison of volatility estimates of S$\&$P 500 series}
    \label{fig:vol_compare_sp}
\end{figure}

The upper plot in Figure \ref{fig:vol_compare_sp} illustrates the logarithm of the return on the closing price for that period, while the lower plot displays the average recorded volatility derived from the Gaussian model, NSVM-1 model, and NSVM-2 model. Notably, the estimated volatility shows a significant rise at points such as $t=60$ and $t=190$, reacting to considerable fluctuations in return values, whether these represent gains or losses. This pattern highlights that the estimated volatility effectively reflects the degree of price movements, indicating the strength of these changes.

\begin{figure}[H]
    \centering
    \includegraphics[width=\linewidth]{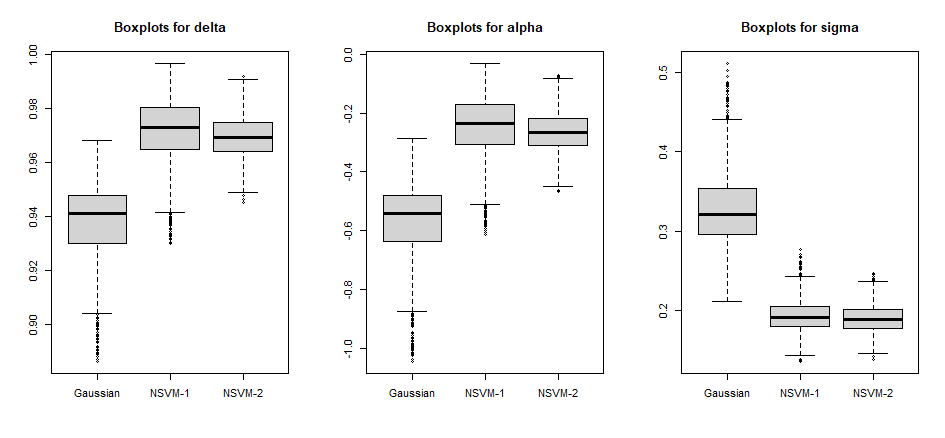}
    \caption{Box-plots of parameter estimates of the SV model, S$\&$P 500 data}
    \label{fig:parameter_est_compare_sp}
\end{figure}

For parameter estimation, we record the means of the posterior distributions, $p(\alpha|h,\delta,\sigma_\nu^2)$, $p(\sigma_\nu^2|h,\alpha,\delta)$, and $p(\delta|h,\alpha,\sigma_\nu^2)$. Figure \ref{fig:parameter_est_compare_sp} displays the box plots of the means based on 100 repetitions of parameter estimates.

In real-world situations, determining the true values of parameters can be quite difficult, unlike simulations where these values are pre-set and easily accessible. This difficulty arises from the intricate nature of actual data, which is frequently affected by a variety of non-financial factors.

\section{Conclusions}

This paper presents a semiparametric method for modeling stochastic volatility (SV). It employs nonparametric estimation for the error distribution to address the limitations of Gaussian assumptions in traditional parametric models. By integrating this nonparametric framework into a Bayesian Markov Chain Monte Carlo (MCMC) approach, we aim to better capture the non-Gaussian features commonly observed in financial return data, enhancing the model's flexibility and accuracy. 

Our extensive simulation findings highlight the benefits of the non-parametric SV models (NSVM-1 and NSVM-2) compared to traditional Gaussian models, especially in the presence of heavy-tailed errors. These non-parametric models consistently exhibit lower bias and variance when evaluated using simulated data across Gaussian, Student-t, and Generalized Error Distributions (GED). Key performance indicators, including the root mean squared error (srMSE), mean absolute error (MAE) and mean absolute percentage error (MAPE), demonstrate that non-parametric models surpass parametric alternatives in all scenarios, with notable improvements in heavy-tailed distributions. These findings suggest that non-parametric models excel in adjusting to non-Gaussian error patterns, resulting in more precise estimates of volatility and parameters.

The empirical examination of S$\&$P 500 data illustrates its practical significance and aligns well with actual volatility measures, like the VIX index. Nevertheless, the key contribution of this study lies in demonstrating the effectiveness of the proposed semiparametric methods through controlled simulations. The findings validate that the NSVM-1 and NSVM-2 models provide a more nuanced representation of financial volatility dynamics, enhancing the precision of risk assessments. 

This paper introduces a novel method for modeling volatility by integrating a semiparametric framework into the SV model. However, the findings have some limitations. Specifically, the model assumes that the error distributions, $u_t$ and $\nu_t$, are independent, which may not reflect reality. Future research could seek to relax this assumption by examining the joint distribution of the error terms and adjusting the likelihood accordingly. Another promising direction is to improve kernel density estimation techniques in this framework, potentially by incorporating dynamic bandwidth selection to better handle time-varying volatility. Additionally, applying this non-parametric approach to multivariate financial models may enhance adaptability and accuracy in capturing intricate financial behaviors across diverse datasets.

\newpage
\bibliography{Reference}

\newpage
\section{Appendix}

\subsection{Effect of sample size in simulation}

We utilized the algorithm with a time series of 500 length and also evaluated it with lengths of 2000 and 5000. These simulations incorporated errors based on Normal, Student's t, and GED distributions. This approach enabled us to evaluate outcomes across different data lengths and assess the algorithm's performance in various contexts.

\begin{figure}[H]
    \centering
    \includegraphics[width=\linewidth]{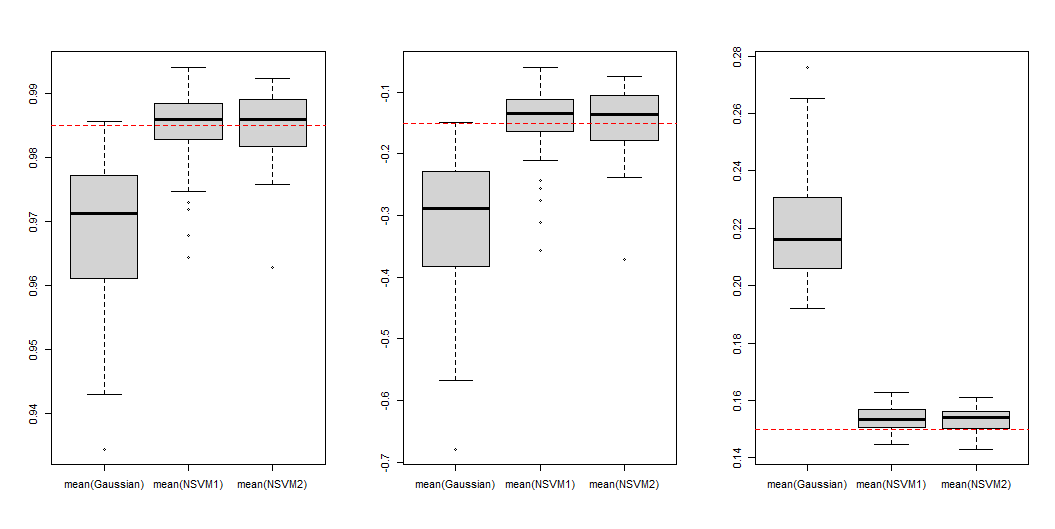}
    \caption{Box-plots of parameter estimates of Gaussian error model, sample size 2000.}
    \label{fig:boxplot_para_est_2000}
\end{figure}

\begin{figure}[H]
    \centering
    \includegraphics[width=\linewidth]{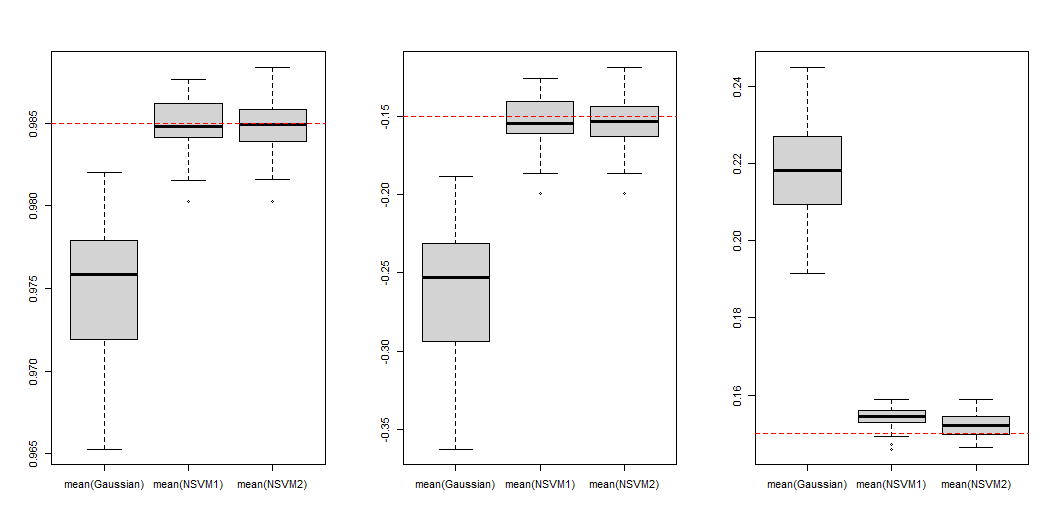}
    \caption{Box-plots of parameter estimates of Student's t error model, sample size 2000.}
    \label{fig:boxplot_para_est_studentt_2000}
\end{figure}

\begin{figure}[H]
    \centering
    \includegraphics[width=\linewidth]{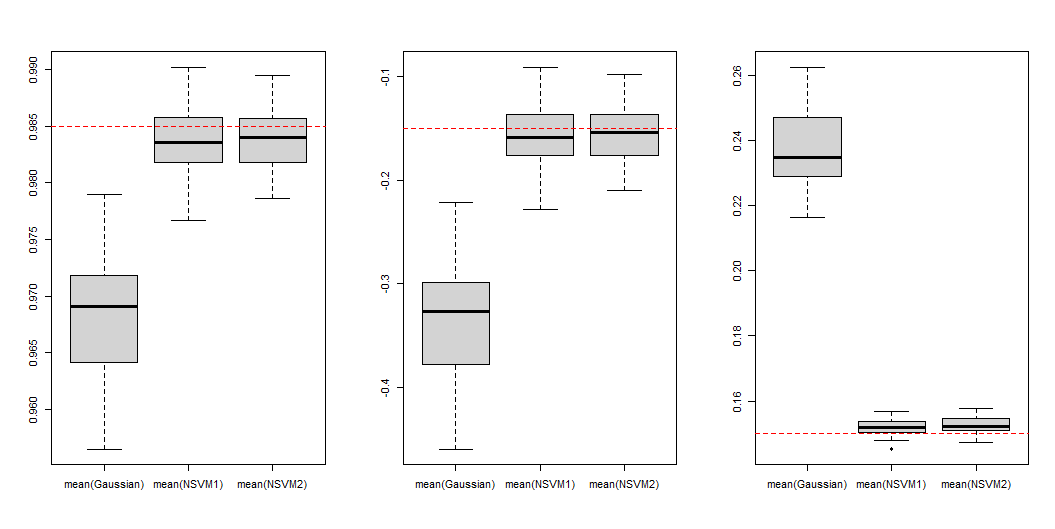}
    \caption{Box-plots of parameter estimates of the GED error model, sample size 2000.}
    \label{fig:boxplot_para_est_ged_2000}
\end{figure}

\begin{figure}[H]
    \centering
    \includegraphics[width=\linewidth]{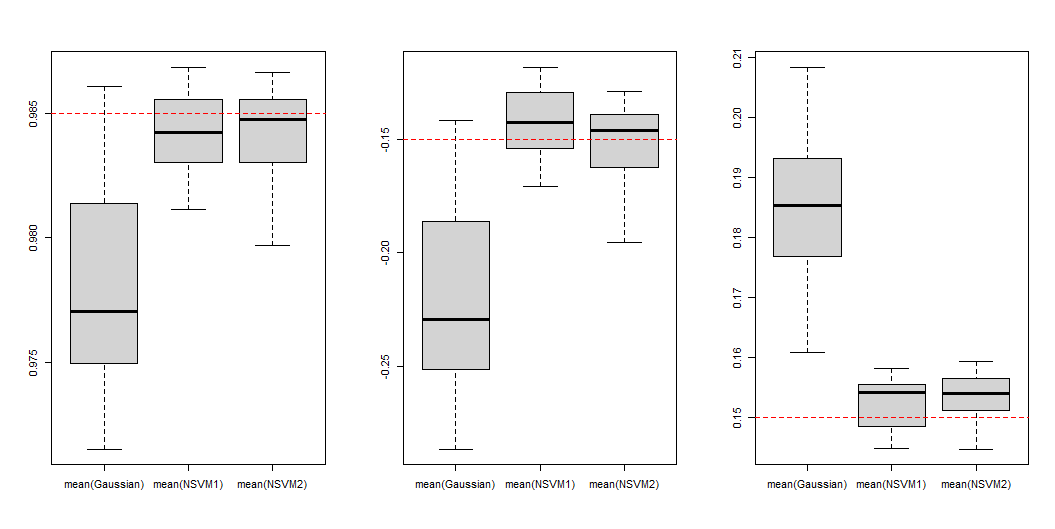}
    \caption{Box-plots of parameter estimates of Gaussian error model, sample size of 5000.}
    \label{fig:boxplot_para_est_5000}
\end{figure}

\begin{figure}[H]
    \centering
    \includegraphics[width=\linewidth]{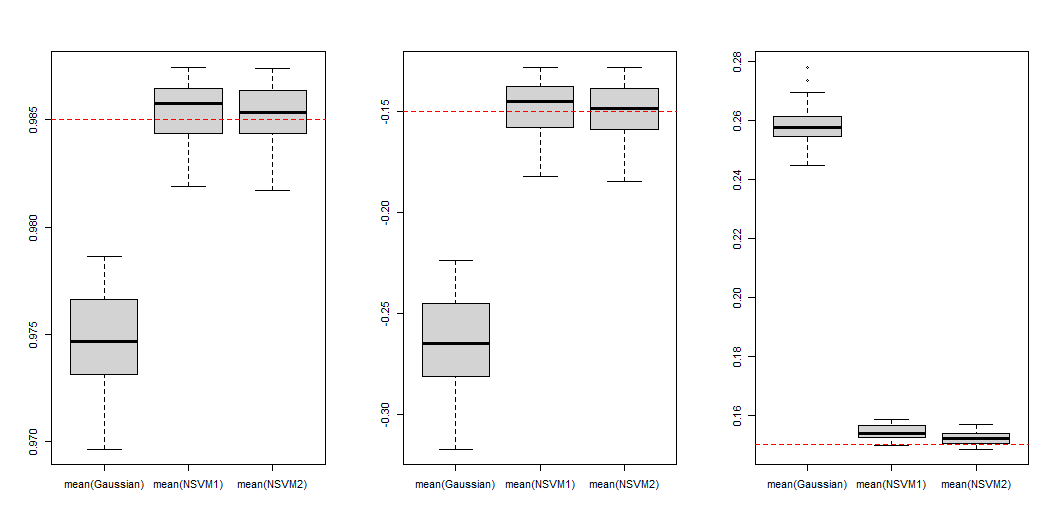}
    \caption{Box-plots of parameter estimates of Student's t error model, sample size 5000.}
    \label{fig:boxplot_para_est_studentt_5000}
\end{figure}

\begin{figure}[H]
    \centering
    \includegraphics[width=\linewidth]{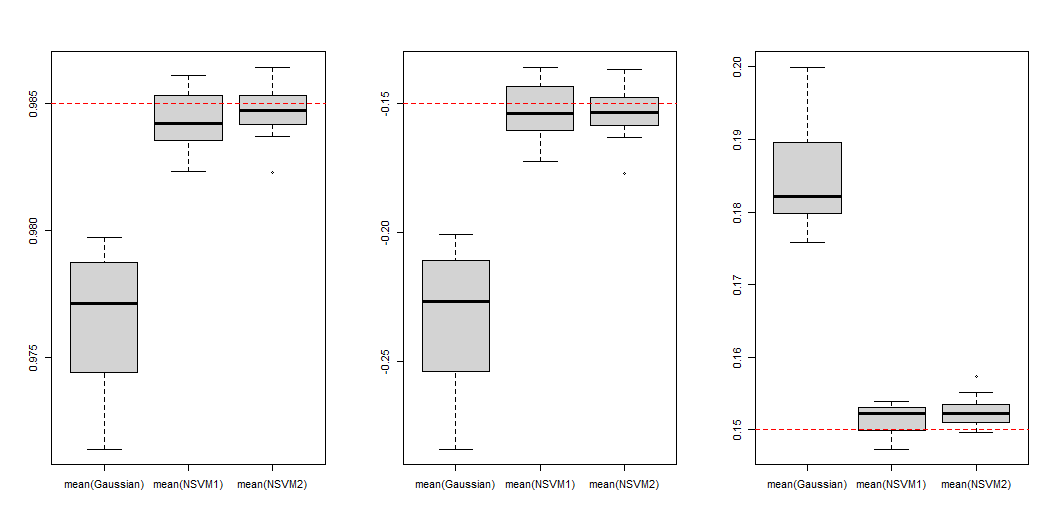}
    \caption{Box-plots of parameter estimates of the GED error model, sample size 5000.}
    \label{fig:boxplot_para_est_ged_5000}
\end{figure}

\subsection{Tuning parameters and bandwidth}

Tuning parameters plays a vital role in performance, ensuring a proper balance between exploration and convergence. The tuning parameter $c^{\star}$ illustrated in equation 
\ref{eq:c_star} must be adjusted with care to achieve optimal outcomes. This adjustment is crucial because an incorrectly set parameter can result in elevated rejection rates and ineffective sampling. We implement a grid search for parameter tuning, adjusting $c^{\star}$ in increments of 0.1 from 0.8 to 1.4. For each value, we fit the model and record the mean of the posterior for the parameter, repeating this process 50 times. Consequently, we have $\bar{\delta_i}$, $\bar{\alpha_i}$ and $\bar{\sigma_{\nu i}}$ for $i = 1, 2, ..., 50$, and the mean square error (MSE) can be defined as $\frac{1}{50}\left( \Sigma \bar{\delta_i} - \delta \right)^2$, $\frac{1}{50}\left( \Sigma \bar{\alpha_i}- \alpha \right)^2$ and $\frac{1}{50} \left(\Sigma \bar{\sigma_{\nu i}}- \sigma_\nu \right)^2$. The results are presented in Table $\ref{table:c_star}$ and Table $\ref{table:c_star_cont1}$. We select $c^\star = 1.2$ for simulations because this choice minimizes the mean squared errors in the estimation of parameters.

\begin{table}[H]    
    \centering
    \resizebox{\textwidth}{!}{
    \begin{tabular}{|cc|c|c|c|c|}
        \hline
        \multicolumn{2}{|c|}{Value of $c^{\star}$} & 0.8 & 0.9 & 1.0 & 1.1  \\ \hline
        \multicolumn{1}{|c|}{\multirow{3}{*}{MSE of $\delta$}}   & Gaussian & $2.567*10^{-4}$ & $2.459*10^{-4}$ & $1.831*10^{-4}$ & $1.263*10^{-4}$   \\ 
        \multicolumn{1}{|c|}{}   & NSVM-1 & $2.131*10^{-4}$ & $1.983*10^{-4}$ & $1.211*10^{-4}$ & $1.067*10^{-4}$ \\ 
        \multicolumn{1}{|c|}{}   & NSVM-2 & $1.862*10^{-4}$ & $1.438*10^{-4}$ & $1.181*10^{-4}$ & $8.856*10^{-5}$ \\ \hline
        \multicolumn{1}{|c|}{\multirow{3}{*}{MSE of $\alpha$}} & Gaussian  & 0.0234 & 0.0261 & 0.0184 & 0.0139  \\  
        \multicolumn{1}{|c|}{} & NSVM-1 & 0.0396 & 0.0376 & 0.0361 & 0.0139  \\
        \multicolumn{1}{|c|}{} & NSVM-2 & 0.0385 & 0.0298 & 0.0236 & 0.0113  \\ \hline
        \multicolumn{1}{|c|}{\multirow{3}{*}{MSE of $\sigma_\nu$}} &  Gaussian & $1.423*10^{-4}$ & $1.127*10^{-4}$ & $5.532*10^{-5}$ & $7.473*10^{-5}$   \\  
        \multicolumn{1}{|c|}{}  &  NSVM-1 & $1.123*10^{-4}$ & $1.142*10^{-4}$ & $7.351*10^{-5}$ & $1.263*10^{-4}$   \\ 
        \multicolumn{1}{|c|}{}   &  NSVM-2  & $2.567*10^{-4}$ & $2.459*10^{-4}$ & $1.831*10^{-4}$ & $9.236*10^{-5}$   \\
        \hline
    \end{tabular}}
    \caption{Table of MSE of parameter estimates by different $c^{\star}$}
    \label{table:c_star}
\end{table}

\begin{table}[H]
    \centering
    \resizebox{\textwidth}{!}{
    \begin{tabular}{|cc|c|c|c|}
        \hline
        \multicolumn{2}{|c|}{Value of $c^{\star}$}  & 1.2 & 1.3 & 1.4 \\ \hline
        \multicolumn{1}{|c|}{\multirow{3}{*}{MSE of $\delta$}}   & Gaussian  & $9.542*10^{-5}$ & $1.434*10^{-4}$ & $1.231*10^{-4}$  \\ 
        \multicolumn{1}{|c|}{}   & NSVM-1 & $1.051*10^{-4}$ & $1.893*10^{-4}$ & $2.231*10^{-4}$\\ 
        \multicolumn{1}{|c|}{}   & NSVM-2 & $9.943*10^{-5}$ & $1.435*10^{-4}$ & $1.641*10^{-4}$\\ \hline
        \multicolumn{1}{|c|}{\multirow{3}{*}{MSE of $\alpha$}} & Gaussian  & 0.0164 & 0.0174 & 0.0189 \\  
        \multicolumn{1}{|c|}{} & NSVM-1  & 0.0135 & 0.0194 & 0.0219 \\
        \multicolumn{1}{|c|}{} & NSVM-2  & 0.0125 & 0.0147 & 0.0253 \\ \hline
        \multicolumn{1}{|c|}{\multirow{3}{*}{MSE of $\sigma_\nu$}} &  Gaussian & $9.348*10^{-5}$ & $1.542*10^{-4}$ & $1.135*10^{-4}$  \\  
        \multicolumn{1}{|c|}{}  &  NSVM-1 & $9.542*10^{-5}$ & $1.434*10^{-4}$ & $1.231*10^{-4}$  \\ 
        \multicolumn{1}{|c|}{}   &  NSVM-2  & $9.853*10^{-5}$ & $1.141*10^{-4}$ & $1.096*10^{-4}$  \\
        \hline
    \end{tabular}}
    \caption{Table of MSE of parameter estimates by different $c^{\star}$ (cont.)}
    \label{table:c_star_cont1}
\end{table}

Selecting the right bandwidth is critical in kernel density estimation; the bandwidth parameter $b$ influences the smoothness of the density estimate by defining the width of the kernel function. In our implementation, we assessed various built-in methods for bandwidth selection available in \textsc{R}, including \texttt{nrd0} \cite{Silverman_kde}, \texttt{bcv} \cite{ScottTerrell1987}, and \texttt{SJ} \cite{SJ}. The results show no noteworthy performance variations among the different bandwidth selection options. As a result, the algorithm adopted the default bandwidth method. \texttt{nrd0}, in \textsc{R}.

\subsection{Detailed algorithms}

This section provides a detailed description of the algorithms discussed in the paper.

\begin{algorithm}
    \begin{algorithmic}
        \Require $Input: (y,v_0,s_0,\delta_0,\sigma_\delta^2,\alpha_0,\sigma_\alpha^2,T,b)$
        \State $N \gets length(y)$
        \State $\delta \gets 1$
        \State $\alpha \gets 0$
        \State $\sigma_\nu^2 \gets 0.1$
        \State $h \gets rep(var(y),N)$
        \State $ln\_h \gets log(h)$
        \State $h\_sample \gets matrix(N,T)$
        \State $\delta\_sample \gets rep(0,T)$
        \State $\alpha\_sample \gets rep(0,T)$
        \State $\sigma_\nu^2\_sample \gets rep(0,T)$
        \For{$ite\ in\ 1\ to\ (T+b)$}
            \For{$i\ in\ 2\ to\ N-1$}
                \State $r \gets sample\_h(y[i],\alpha,\delta,\sigma_\nu^2,ln\_h[i-1],ln\_h[i+1])$
                \State $h[i] \gets r\$ h\_new$
                \State $ln\_h \gets log(h[i])$
            \EndFor
            \State $ln\_h[1] \gets rnorm(1,\alpha+\delta*ln\_h[2],\sqrt{\sigma_\nu^2})$
            \State $ln\_h[N] \gets rnorm(1,\alpha+\delta*ln\_h[N-1],\sqrt{\sigma_\nu^2})$
            \State $h[1] \gets exp(ln\_h[1])$
            \State $h[N] \gets exp(ln\_h[N])$
            \State $s_1 \gets sum(ln\_h)$
            \State $s_2 \gets sum(ln\_h^2)$
            \State $s_3 \gets sum(ln\_h[1:(N-1)]*ln\_h[2:N])$
            \State $s' \gets s_0+(N-1)\alpha^2+(1+\delta^2)s_2-\delta^2(ln\_h[N])^2-ln\_h[1]^2-2\alpha((1-\delta)s_1-ln\_h[1]+\delta ln\_h[N])-2\delta s_3$
            \State $\sigma_\nu^2 \gets sample\_\sigma_\nu^2(v_0,s',\alpha,\delta,h,\sigma_\nu^2\_sample[ite-1])$
            \State $\alpha \gets sample\_\alpha(\alpha_0,\sigma_\alpha^2,\delta,\sigma_\nu^2,h,\alpha\_sample[ite-1])$
            \State $\delta \gets sample\_\delta(\delta_0,\sigma_\delta^2,\alpha,\sigma_\nu^2,h,\delta\_sample[ite-1])$
            \If{$ite>b$}
                \State $\alpha\_sample[ite-b] \gets \alpha$
                \State $\delta\_sample[ite-b] \gets \delta$
                \State $\sigma_\nu^2\_sample[ite-b] \gets \sigma_\nu^2$
                \State $h\_sample[ite-b] \gets h$
            \EndIf
        \EndFor
        \State \Return $(h\_sample,\alpha\_sample,\delta\_sample,\sigma_\nu^2\_sample)$
    \end{algorithmic}
    \caption{Main}
    \label{Alg:sampler_main}
\end{algorithm}

\begin{algorithm}
    \begin{algorithmic}
        \Require $Input: (y,\alpha,\delta,\sigma_\nu^2,ln\_h_{i-1},ln\_h_{i+1})$
        \State $\mu \gets \frac{\delta (ln\_h_{i-1}+ln\_h_{i+1})+(1+\delta)\alpha}{1+\delta^2}$
        \State $\sigma^2 \gets \frac{\sigma_\nu^2}{1+\delta^2}$
        \State $\lambda \gets \frac{1-2exp(\sigma^2)}{1-exp(\sigma^2)}+0.5$
        \State $\phi \gets (\lambda-1)exp(\mu+\frac{\sigma^2}{2})+\frac{y^2}{2}$
        \State $qmode = \frac{\phi}{1+\lambda}$
        \State $ln\_c \gets ln(1.2)+ln\_p(y,qmode,\alpha,\delta,\sigma_\nu^2)-ln(dinvgamma(qmode,\lambda,\phi))$
        \State $h\_new \gets rinvgamma(1,\lambda,\phi)$
        \State $threshold_1 \gets \frac{exp(ln\_p(y,h\_new,qmode,\alpha,\delta,\sigma_\nu^2)-ln\_c)}{dinvgamma(h\_new,\lambda,\phi)}$
        \State $roll \gets runif(1)$
        \While{$roll>threshold_1$}
            \State $h\_new \gets rinvgamma(1,\lambda,\phi)$
            \State $threshold_1 \gets \frac{exp(ln\_p(y,h\_new,qmode,\alpha,\delta,\sigma_\nu^2)-ln\_c)}{dinvgamma(h\_new,\lambda,\phi)}$
            \State $roll \gets runif(1)$
        \EndWhile
        \If{$threshold_1 \geq 1$}
            \State $threshold_2 \gets exp(ln\_p(y,h\_new,qmode,\alpha,\delta,\sigma_\nu^2)-ln\_p(y,h,qmode,\alpha,\delta,\sigma_\nu^2)+log(dinvgamma(h,\lambda,\phi))-log(dinvgamma(h\_new,\lambda,\phi)))$
            \State $roll \gets runif(1)$  
            \If{$roll>threshold_2$}
                \State $h\_new \gets h$
            \EndIf
        \EndIf
        \State \Return $h\_new$
    \end{algorithmic}
    \caption{Sampler for $h$}
    \label{alg:sampler_h}
\end{algorithm}

\begin{algorithm}
    \begin{algorithmic}
        \Require $Input: (v_0,s',\alpha,\delta,h,\sigma^2_{left})$
        \State $q(\sigma_\nu^2) \sim IG(\frac{v_0+N-1}{2},\frac{s'}{2})$
        \State $p(\sigma_\nu^2) \gets p(\sigma_\nu^2|h,\alpha,\delta)$
        \State $c \gets \frac{p(\sigma_\nu^2)}{q(\sigma_\nu^2)}$, at $\sigma_\nu^2$ = mode of $q(\sigma_\nu^2)$, i.e. $\frac{s'/2}{\frac{v_0+N-1}{2}+1}$
        \State $ln\_c \gets log(p(\sigma_\nu^2))-log(q(\sigma_\nu^2))$
        \State $\sigma^2_{new} \gets rinvgamma(1,\frac{v_0+N-1}{2},\frac{s'}{2})$
        \State $threshold_1 \gets \frac{exp(p(\sigma_\nu^2)-ln\_c)}{dinvgamma(\sigma^2_{new},\frac{v_0+N-1}{2},\frac{s'}{2})}$
        \State $roll \gets runif(1)$
        \While{$roll>threshold_1$}
            \State $\sigma^2_{new} \gets rinvgamma(1,\frac{v_0+N-1}{2},\frac{s'}{2})$
            \State $threshold_1 \gets \frac{exp(p(\sigma_\nu^2)-ln\_c)}{dinvgamma(\sigma^2_{new},\frac{v_0+N-1}{2},\frac{s'}{2})}$
            \State $roll \gets runif(1)$
        \EndWhile
        \If{$threshold_1 \geq 1$}
            \State $threshold_2 \gets \frac{p(\sigma^2_{new})/p(\sigma^2_{left})}{dinvgamma(\sigma^2_{new},\frac{v_0+N-1}{2},\frac{s'}{2})/dinvgamma(\sigma^2_{left},\frac{v_0+N-1}{2},\frac{s'}{2})}$
            \State $roll \gets runif(1)$  
            \If{$roll>threshold_2$}
                \State $\sigma^2_{new} \gets \sigma^2_{left}$
            \EndIf
        \EndIf
        \State \Return $\sigma^2_{new}$
    \end{algorithmic}
    \caption{Sampler for $\sigma_\nu^2$}
    \label{alg:sampler_sigma}
\end{algorithm}

\begin{algorithm}
    \begin{algorithmic}
        \Require $Input: (\delta_0,\sigma_\delta^2,\alpha,\sigma_\nu^2,h,\delta_{left})$
        \State $qfunc\_mean\_\delta \gets \frac{\sigma_\nu^2\delta_0+\sigma_\delta^2(s_3-\alpha(S_1-\ln h_N))}{\sigma_\nu^2+\sigma_\delta^2(s_2-(\ln h_N)^2)}$
        \State $qfunc\_var\_\delta \gets \frac{\sigma_\nu^2\sigma_\delta^2}{\sigma_\nu^2+\sigma_\delta^2(s_2-(\ln h_N)^2)}$
        \State $q(\delta) \sim N\left(qfunc\_mean\_\delta,qfunc\_var\_\delta\right)$
        \State $p(\delta) \gets p(\delta|h,\alpha,\sigma_\nu^2)$
        \State $c \gets \frac{p(\delta)}{q(\delta)}$, at $\delta$ = mode of $q(\delta)$
        \State $ln\_c \gets log(p(\delta))-log(q(\delta))$
        \State $\delta_{new} \gets rnorm(1,qfunc\_mean\_\delta,qfunc\_var\_\delta)$
        \State $threshold_1 \gets \frac{exp(p(\delta)-ln\_c)}{dnorm(\delta\_new,qfunc\_mean\_\delta,qfunc\_var\_\delta)}$
        \State $roll \gets runif(1)$
        \While{$roll>threshold_1$}
            \State $\delta\_new \gets rnorm(1,qfunc\_mean\_\delta,qfunc\_var\_\delta)$
            \State $threshold_1 \gets \frac{exp(p(\delta)-ln\_c)}{dnorm(\delta_{new},qfunc\_mean\_\delta,qfunc\_var\_\delta)}$
            \State $roll \gets runif(1)$
        \EndWhile
        \If{$threshold_1 \geq 1$}
            \State $threshold_2 \gets \frac{p(\delta_{new})/p(\delta_{left})}{dnorm(\delta_{new},qfunc\_mean\_\delta,qfunc\_var\_\delta)/dnorm(\delta_{left},qfunc\_mean\_\delta,qfunc\_var\_\delta)}$
            \State $roll \gets runif(1)$  
            \If{$roll>threshold_2$}
                \State $\delta_{new} \gets \delta_{left}$
            \EndIf
        \EndIf
        \State \Return $\delta_{new}$
    \end{algorithmic}
    \caption{Sampler for $\delta$}
    \label{alg:sampler_delta}
\end{algorithm}

\begin{algorithm}
    \begin{algorithmic}
        \Require $Input: (\alpha_0,\sigma_\alpha^2,\delta,\sigma_\nu^2,h,\alpha_{left})$
        \State $qfunc\_mean\_\alpha \gets \frac{\sigma_\alpha^2((1-\delta)s_1-\ln h_1+\delta\ln h_N)+\sigma_\nu^2\alpha_0}{\sigma_\nu^2+(N-1)\sigma_\alpha^2}$
        \State $qfunc\_var\_\alpha \gets \frac{\sigma_\nu^2\sigma_\alpha^2}{\sigma_\nu^2+(N-1)\sigma_\alpha^2}$
        \State $q(\alpha) \sim N\left(qfunc\_mean\_\alpha,qfunc\_var\_\alpha\right)$
        \State $p(\alpha) \gets p(\alpha|h,\delta,\sigma_\nu^2)$
        \State $c \gets \frac{p(\alpha)}{q(\alpha)}$, at $\alpha$ = mode of $q(\alpha)$
        \State $ln\_c \gets log(p(\alpha))-log(q(\alpha))$
        \State $\alpha_{new} \gets rnorm(1,qfunc\_mean\_\alpha,qfunc\_var\_\alpha)$
        \State $threshold_1 \gets \frac{exp(p(\alpha)-ln\_c)}{dnorm(\alpha_{new},qfunc\_mean\_\alpha,qfunc\_var\_\alpha)}$
        \State $roll \gets runif(1)$
        \While{$roll>threshold_1$}
            \State $\alpha_{new} \gets rnorm(1,qfunc\_mean\_\alpha,qfunc\_var\_\alpha)$
            \State $threshold_1 \gets \frac{exp(p(\alpha)-ln\_c)}{dnorm(\alpha_{new},qfunc\_mean\_\alpha,qfunc\_var\_\alpha)}$
            \State $roll \gets runif(1)$
        \EndWhile
        \If{$threshold_1 \geq 1$}
            \State $threshold_2 \gets \frac{p(\alpha_{new})/p(\alpha_{left})}{dnorm(\alpha_{new},qfunc\_mean\_\alpha,qfunc\_var\_\alpha)/dnorm(\alpha_{left},qfunc\_mean\_\alpha,qfunc\_var\_\alpha)}$
            \State $roll \gets runif(1)$  
            \If{$roll>threshold_2$}
                \State $\alpha_{new} \gets \alpha_{left}$
            \EndIf
        \EndIf
        \State \Return $\alpha_{new}$
    \end{algorithmic}
    \caption{Sampler for $\alpha$}
    \label{alg:sampler_alpha}
\end{algorithm}

\end{document}